\documentclass[runningheads]{svjour2}

\usepackage[left=4cm,top=4.19cm,right=4cm,bottom=4.19cm]{geometry}

\usepackage[english]{babel}

\usepackage[T1]{fontenc}
\usepackage{mathrsfs}
\usepackage{graphicx}
\usepackage{subfigure}
\usepackage{amsmath}
\usepackage{amssymb}
\usepackage{wasysym} 

\usepackage{microtype}

\newcommand{\ie}{i.e.,\ }
\newcommand{\eg}{e.g.,\ }

\newcommand{\beq}{\begin{equation}}
\newcommand{\eeq}{\end{equation}}
\newcommand{\rmd}{\mathrm{d}}

\newcommand{\qder}[2]{\operatorname{\mathscr{D}}^{#1}_{#2}}

\begin{document}

\title{$q$-deformed Loewner evolution}
\date{}

\author{Marco Gherardi \and Alessandro Nigro}
\authorrunning{M.\hspace{0.2em}Gherardi \and A.\hspace{0.2em}Nigro}

\institute{M.\ Gherardi \and A.\ Nigro \at Dipartimento di Fisica, Universit\`a degli Studi di Milano, via Celoria 16, 20133 Milano, Italy
	\and Istituto Nazionale di Fisica Nucleare, sezione di Milano, via Celoria 16, 20133 Milano, Italy\\
	\email{Marco.Gherardi@mi.infn.it}}

\maketitle

\vspace{-0.5cm}
\begin{abstract}
The Loewner equation, in its stochastic incarnation introduced by Schramm,
is an insightful method for the description of critical random curves
and interfaces in two-dimensional statistical mechanics.
Two features are crucial, namely conformal invariance and a conformal version
of the Markov property.
Extensions of the equation have been explored in various directions, in order to
expand the reach of such a powerful method.
We propose a new generalization based on $q$-calculus,
a concept rooted in quantum geometry and non-extensive thermodynamics;
the main motivation is the explicit breaking of the Markov property, while
retaining scale invariance in the stochastic version.
We focus on the deterministic equation and give some exact solutions;
the formalism naturally gives rise to multiple mutually-intersecting curves.
A general method of simulation is constructed --- which can be easily extended
to other $q$-deformed equations --- and is applied to both the deterministic
and the stochastic realms.
The way the $q\neq 1$ picture converges to the classical one is explored as well.
\end{abstract}


\section{Introduction}

Schramm-Loewner evolution (SLE) is a stochastic process 
whose values are conformal mappings on a domain in the complex plane.
It was proposed by Schramm in 1999 \cite{Schramm:2000} as a stochastic version
of an equation studied by Loewner at the beginning of the twentieth century
in connection with the Bieberbach conjecture.
The core idea is to describe a family of growing domains
by means of the conformal maps (called uniformizing) 
that send the complement of the domains 
onto the whole space again (the disc or the upper half plane, for instance);
this is what one usually does in order to simplify
the computation of the electrostatic potential with conductors of difficult shapes.
The time evolution of these domains is then captured by
a first-order differential equation for the uniformizing maps.
If the domain is a growing simple (or non-self-traversing) curve then
the equation contains a real function of time, called driving function, which completely
characterizes the curve.
Choosing a stochastic driving function then gives a statistical ensemble of curves.
As Schramm showed, requiring that the latter be conformally invariant and
Markovian essentially fixes the stochastic process to Brownian motion of
arbitrary diffusivity.
Here the term Markovian is not used strictly in the usual sense: it roughly means that
the growth of the curve at time $t$ does not depend on the full
history at all times $\tau\leq t$, but only on the uniformizing map
at time $t$.
We will briefly introduce the relevant definitions in the next section; the reader interested
in a wider introduction to SLE will find many authoritative reviews
(see for instance \cite{Cardy:SLEreview,KagerNienhuis:SLEreview,Gruzberg:SLEreview}, 
aiming at physicists, and \cite{Lawler:book}, aiming at mathematicians).

As the diffusivity of the Brownian driving function varies,
a continuum of universality classes is obtained, and
correspondences to numerous models in statistical mechanics
have been discovered, including the Ising model \cite{Duminil-CopinSmirnov:2011}
(even in three dimensions \cite{SaberiDashti-Naserabadi:2010}),
spin glasses \cite{AmorusoHartmann:2006},
the Ashkin-Teller model \cite{IkhlefRajabpour:2012},
the Potts model \cite{GliozziRajabpour:2010},
walks and polymers on the lattice \cite{Schramm:2000,Kennedy:2002prl,Gherardi:2009,Daryaei:2012},
non-equilibrium phenomena \cite{Boffetta:2006,Nezhadhaghighi:2011}
(even in three dimensions \cite{Thalabard:2011}),
and the Gaussian free field \cite{ShrammSheffield:2009}.
A deep connection to Conformal Field Theory (CFT) has been unveiled
\cite{BauerBernard:2003}, as well as with logarithmic CFT 
\cite{SaintAubin:2009,Rasmussen:2004},
and some generalizations of SLE have been based on it
\cite{LesageRasmussen:2004,BettelheimGruzberg:2005,Nazarov:2012,MoghimiAraghiRajabpourRouhani:2004}.
On the other hand, various generalizations and modifications have been proposed
by extending or changing the driving function,
for instance by using L\'evy flights \cite{RushkinOikonomouKadanoffGruzberg:2006},
or discrete-scale invariant functions \cite{Nezhadhaghighi:2010}.
Of course, not all statistical mechanics models on the lattice naturally
lead to curves whose continuum limit should be SLE
(\eg negative-weight percolation \cite{NorrenbrockHartmann:2012},
disordered solid-on-solid models \cite{SchwarzKarrenbauer:2009},
bimodal Edwards-Anderson spin glass \cite{Risau-Gusman:2008});
while conformal invariance is a natural requirement
for models at criticality, the Markov property is not to be
expected in general.

Our aim in this paper is to propose and study a generalization
of the Loewner equation (both deterministic and stochastic) which
is based on a widespread and natural concept, that of $q$-deformation.
Instead of changing the driving function or the form of the equation,
which is the usual strategy for extending SLE, we rely on a
redefinition of the derivative operator, inspired by the methods
of $q$-calculus.
The main rationale behind this choice is the wish to explicitly
break the Markov property, while preserving the scale invariance,
which is expected for critical phenomena
(massive extensions of SLE, breaking scale and conformal invariance,
have been studied \cite{MakarovSmirnov:2010}).
The definition of conformal invariance usually employed in this context
is actually dependent on the Markov property, so we do not expect
to retain it in its full form.
What is left is a weaker version, which is essentially the combination
of scale invariance and the conformal transport principle
(a way of defining how to transport measures between conformally
equivalent domains).

The history of $q$-calculus started more than two hundred years ago
with Euler's work on partitions of integers, where a generating
function of the form (\ref{eq:pochhammer}) appears, but only in
the nineteenth century did it gain momentum with the work of
Heine on the generalization of hypergeometric series. 
It was F.~H.~Jackson, in the early twentieth century, who first introduced
and studied systematically  the $q$-derivative operator 
that we will use in our extension of the Loewner equation.
This type of deformed calculus is sometimes called \emph{quantum calculus}.
The reasons for the use of the word ``quantum'' are connected to the
natural appearance of $q$-differential operators in the theory of
the so-called \emph{quantum groups} (the name is due to Drinfeld),
whose importance in physics has grown in the past decades.
For an overview, see \cite{CastellaniWess:1996}
and \cite{Bonatsos:1999}; for a very good and concise
introduction to $q$-calculus and applications to number theory
and combinatorics, see \cite{KacCheung:2002}.

We concentrate here mainly on the deterministic Loewner equation,
which already \hyphenation{pres-ents} presents a rich and challenging behavior,
but we expect its stochastic version to be more directly
relevant for applications in statistical physics.
In particular, 
a form of $q$-calculus is involved in the description of
non-extensive statistical mechanics \cite{Tsallis:2002},
whose scope of application ranges from phenomena where the Loewner 
approach has already proved useful, like the Ising model
\cite{SilvaStanley:1996} and viscous fingering \cite{Grosfils:2006}, 
to broader and more general subjects, as generalizations of the $\alpha$-stable
distributions \cite{UmarovTsallis:2010} and anomalous diffusion \cite{Tsallis:2005}.
We do not pursue these connections in this paper, but they
constitute the physical motivations for a systematic approach to
$q$-deformed equations in statistical mechanics.

In section \ref{section:definition} we recall the basic concepts
of $q$-calculus and of the theory of (Schramm-)Loewner evolutions
that are needed, and we define the extension of the Loewner equation 
that we propose.
Its main properties will be discussed in section \ref{section:properties},
namely scale invariance and the (lack of) Markov property.
In section \ref{section:special_cases} we derive an exact solution
to the equation in a special case (constant driving function), 
and also discuss the critical case corresponding to the average
behavior of Brownian motion.
Then we work our way through the definition of a feasible
algorithm in section \ref{section:algorithm}, and apply it to the numerical
study of the equation in section \ref{section:numerical_results}.
We find that the equation defines in general an infinite number of curves, 
growing in the upper half plane. Contrary to the multiple SLEs usually studied
\cite{BauerBernardKytola:2005}, these curves can intersect.
We focus on the multiple lines which are obtained for the critical deterministic
driving function (the square root), and study the complex zero-pole pattern that emerges,
which allows to appreciate the non-trivial way in which this realization of a $q$-deformation
converges to the $q=1$ classical physics.

\section{\label{section:definition}Background and definition}

Let us first gather the main ingredients of $q$-calculus
that will be of use in our context.
The $q$-derivative operator is defined as
\beq\label{eq:qder}
\qder{q}{x}h(x) = \frac{h(qx)-h(x)}{(q-1)x}.
\eeq
The parameter $q$ can be a complex number in the most general setting,
but we will restrict to $q\in(0,1)$ in the following, unless explicitly stated.
Note that in the limit $q\to 1$, which we will call \emph{classical},
the $q$-derivative reduces to the ordinary derivative.
It is immediately verified that $\qder{q}{}$ is indeed a linear operator,
when acting on a vector space of functions.
An analogue of the Leibniz rule holds:
\beq
\qder{q}{x}\left[f(x)g(x)\right]=f(qx)\qder{q}{x}g(x)+\qder{q}{x}f(x)g(x).
\eeq 
On the other hand, the chain rule is more problematic, and the general
formula requires two derivatives with different values of the increment $q$ \cite{Larsson:2003};
nonetheless, the classical rule holds when the composed function is 
linear homogeneous; in fact,
\beq
\qder{q}{x}f(cx)=c\,\qder{q}{cx}f(cx).
\eeq
Geometric sums are widespread in $q$-calculus,
so that a special symbol is reserved to them
(other symbols, such as curly brackets, are sometimes found in the literature):
\beq\label{eq:qnumber}
\left[n\right]_q=\sum_{j=1}^n q^{j-1}=\frac{1-q^n}{1-q};
\eeq
this is called the $q$-analogue of $n$, or the $q$-number.
It is encountered for instance in the $q$-derivative of the power law,
\beq\label{eq:powerlawderivative}
\qder{q}{x} x^n = \left[n\right]_q x^{n-1},
\eeq
which can be directly checked from the definition.
We will extend (\ref{eq:qnumber}) to non-integer real constants $\alpha$,
and simply write $\left[\alpha\right]_q$ for $(1-q^\alpha)/(1-q)$;
the rule (\ref{eq:powerlawderivative}) holds for non-integer exponents as well.
The factorial $\left[n\right]_q!$ is defined as $1$ if $n=0$, and as the product of all the $q$-numbers
from $\left[1\right]_q$ to $\left[n\right]_q$ otherwise.
The inverse operation of $q$-derivation has been introduced by
Jackson in the early days of the development of $q$-calculus.
It can be shown that
\beq\label{eq:jacksonintegral}
F(x)=(1-q)x\sum_{j=0}^\infty q^j f(q^j x)   
\eeq
is the unique function (up to a function with null $q$-derivative) whose $q$-derivative is $f(x)$;
the expression in the right-hand side is sometimes called \emph{Jackson integral}.

Let us now briefly recall the ordinary Loewner equation and SLE.
We will focus on the standard \emph{chordal} process, \ie a measure on
curves starting from the origin and growing towards infinity
in the upper half plane $\mathbb{H}$.
Let $U_t$ be a continuous real function of time,
which will be called \emph{driving function}.
Let $g_t(z)$ be the solution to the Loewner equation:
\beq\label{eq:loewner}
\frac{\rmd}{\rmd t}g_t(z)=\frac{2}{g_t(z)-U_t}
\eeq
with initial condition $g_0(z)=z$ (note that the point $z$ only enters
the equation through the initial condition).
If at some time the denominator becomes zero, then the solution does not
exist from that time on.
The set of points $z$ for which the solution does not exist at time $t$ is called \emph{hull}
and is usually denoted by $K_t$.
Then it can be shown \cite{Lawler:book} that $g_t(z)$ is the (unique) conformal map
of $\mathbb{H}\setminus K_t$ onto $\mathbb{H}$ that maps 
$z=\infty$ to itself, and whose derivative at infinity is $1$.
The solution $g_t(z)$ is called the \emph{uniformizing map} of the hull $K_t$.
In other words, by solving the Loewner equation for each starting point $z$ up
to time $t$, one obtains a map in $z$ whose domain is the complement of
a hull $K_t$.
It is clear from the definition that $K_t$ is a \emph{growing} family of hulls,
meaning that $K_s\subseteq K_t$ if $s<t$.
The parametrization is fixed by the form of the equation; it could be changed
by changing the numerator in (\ref{eq:loewner}).

If the driving function is a stochastic process, then the measure on its
realizations becomes (by means of the Loewner equation) a measure
on the growing hulls.
In particular, the work initiated by Schramm has shown that if $U_t$
is chosen to be the rescaled Brownian motion $\sqrt{\kappa}B_t$
then the hulls are fractal curves, whose properties
depend on $\kappa$ (for instance, their fractal dimension
is $1+\kappa/8$).
This stochastic version of the Loewner equation is chordal SLE.
It is defined naturally on the upper half plane, but it can
be transported onto a domain $\mathbb D$ by means of
a conformal map $\phi:\mathbb{H}\to\mathbb{D}$, by defining
$\tilde g_t=\phi\circ g_t\circ\phi^{-1}$.
This is the principle of \emph{conformal transport}.
Independently of $\kappa$, SLE curves satisfy two basic
properties (see for instance \cite{Cardy:SLEreview}):
\begin{itemize}
\item (Conformal invariance) If  $\phi$ is a conformal mapping of
the interior of a domain $\mathbb{D}$ onto the interior of a domain $\mathbb{D'}$,
such that $\phi(a)=a'$ and $\phi(b)=b'$, then the measure  on the curves
from $a$ to $b$ in $\mathbb{D}$ is the same as the measure on the curves
from $a'$ to $b'$ in $\mathbb{D'}$.
\item (Markov property) Denote a generic curve by $\gamma$ and
split it into two pieces, $\gamma_1$ and $\gamma_2$, the first
curve going from point $a$ to point $a'$ and the second curve going
from $a'$ to $b$.
Then the measure on the curves from $a$ to $b$ in $\mathbb{D}$
conditioned to start with $\gamma_1$ is the same as the measure
on the curves from $a'$ to $b$ in $\mathbb{D}\setminus\gamma_1$.
\end{itemize}
Thanks to conformal transport (\ie the definition of $\tilde g_t$ given above),
any measure in a domain $\mathbb{D}$ can be transported to any other domain
$\mathbb{D'}$, so that conformal invariance seems a rather tautological
property. In fact, it becomes meaningful when a prescription for comparing
the measures in different domains is given, for instance via the continuum limits
of critical lattice models.
Without relying on the latter, the strategy employed for SLE --- and historically
one of the seminal observations by Schramm --- is to let $\mathbb{D'}$ be
related to $\mathbb{D}$ as in the Markov property ($\mathbb{D'}=\mathbb{D}\setminus\gamma_1$).
But since we are seeking a generalization where the Markov property is broken,
the only non-trivial choice left is $\mathbb{D'}=\mathbb{D}$,
so we are interested in invariance under automorphisms of $\mathbb{D}$
(see the discussion in Sec.~\ref{section:properties}).

We are now in the position to
introduce a generalization of the Loewner equation based on $q$-calculus,
which will be called $q$-LE (or \emph{$q$-deformed Loewner evolution})
in its deterministic version and $q$-SLE in its stochastic version.
We consider the following functional equation:
\beq\label{eq:qsle}
\qder{q}{t}g_t(z)=\frac{2}{g_{\sqrt{q}t}(z)} - \qder{q}{t}\xi_t,\quad\quad \lim_{t\to 0^+}g_t(z)=z,
\eeq
where $\xi_t$ is the driving function
(ordinary Brownian motion rescaled by $\sqrt{\kappa}$ for $q$-SLE),
and $z$ is a point of the upper half plane $\mathbb H$.
In the following we fix $q\in(0,1)$, unless otherwise specified.
Notice that the q-derivative of the Brownian motion is well-defined,
since no limit is present.
This is one reason why we prefer to write the equation in this way, instead of
the more common form which holds for $\hat g_t=g_t+\xi_t$.

Given a first-order differential equation in normal form, 
its deformation is not in general uniquely defined.
Once a $q$-derivative is substituted for the ordinary derivative in the
left-hand side, one is faced with the arbitrariness of choosing where
to evaluate the right-hand side, since every function of $q$, provided that
its limit for $q\to 1$ is $t$, will do the trick.
We choose $\sqrt{q}t$ for symmetry reasons.
First of all, this is the geometric mean between the two extremes
$t$ and $qt$, and a duality $q\mapsto 1/q$ can then be found (see Sec.~\ref{section:properties}).
Moreover, such a choice gives rise to a special solution for
the constant-driving case which is a natural generalization of the classical one
(see Sec.~\ref{section:qslit}).

An observation is in order regarding the meaning of an equation such as (\ref{eq:qsle}).
It could be objected that it is equivalent to solving the original equation iteratively,
simply being a discrete version of it, although with a $q$-derivative instead of
the more natural (in this context) $h$-derivative (see Bauer's work on
the discrete Loewner evolution \cite{Bauer:2003}).
The difference is that in the present approach we wish to find
solutions $g_t$ which satisfy the functional equation \emph{for every} $t$,
not just on a discrete set of times $\left\{t_i\right\}$.
This is similar to the usual procedure of extending
a function defined on the natural numbers (\eg the factorial)
to the real or complex field (\eg the gamma function).
Owing to the linearity of the $q$-difference operator, 
if $g_t$ is a solution to a $q$-deformed equation such as (\ref{eq:qsle}),
any $q$-constant function $h_t$ can be added to $g_t$ to yield another solution.
The general constants in $q$-calculus are the $q$-periodic functions, 
\ie functions $h_t$ such that $h_{qt}=h_t$ for all $t$.
It is easy to see that no such function, beyond ordinary constants,
has a limit as $t\to 0$.
This is the reason why in the definition (\ref{eq:qsle}) we specify the initial condition $g_0(z)=z$
through a continuity condition in $t=0$.
Equivalently, one can write the Jackson integral
[see Eq.~(\ref{eq:gjackson}) in Sec.~\ref{section:algorithm}] and state that,
provided the integral converges, it gives the unique primitive
that is continuous in $0$ (see Proposition 18.1 in \cite{KacCheung:2002}).

Note that ordinary Loewner equations give rise to conformal
maps on $\mathbb H$, thanks to the properties of ordinary differential equations
(analiticity in $z$) and the particular form of the equation (uniqueness).
We should not expect to find such strong properties in the
weaker $q$-deformed case; in particular, we anticipate that
the maps will not be 1-to-1, and their range will not be
restricted to $\mathbb H$.
However, the process is symmetric with respect to the real axis
(see Sec.~\ref{section:properties}), so one can restrict to the half plane nonetheless.

We now need to define the central object of interest in the
framework of the Loewner equation: the growing hull.
In accordance with the definition for the ordinary case,
we define the hull $K_t$ grown at time $t$ as the complement in
$\mathbb{H}$ of the set of all points $z$
for which the right hand side of (\ref{eq:qsle}) has
not become singular up to time $t$,
\beq
\label{eq:hull}
K_t=\left\{z\in\mathbb{H}:g_{\sqrt{q}\tau}(z)=0, \tau\leq t\right\}.
\eeq
For ordinary Loewner equations, $K_t$ grows ``one point at a time'',
\ie a single point is mapped to $z=0$ at each $t$; 
roughly speaking, this is a consequence
of the invertibility of $g_t$. In the deformed case, however,
we expect the hull to grow in a more general non-local way.

\section{\label{section:properties}Properties}

Let us first state a duality property, which justifies the choice we made
of evaluating the right-hand side of the Loewner equation in $\sqrt{q}t$.
Writing (\ref{eq:qsle}) at time $t=q^{-1}\tau$ yields
\beq\label{eq:duality}
\qder{1/q}{\tau}g_{\tau}(z)=\frac{2}{g_{\sqrt{1/q}\,\tau}(z)}-\qder{1/q}{\tau}\xi_\tau.
\eeq
Equation (\ref{eq:duality}) allows one to extend any definition and result valid for $q<1$
to $q>1$.
The classical case $q=1$ is the self-dual point.

Let us now consider the properties of the stochastic equation, in particular
when $\xi_t$ is (rescaled) Brownian motion.
The reasons for our interest in studying the $q$-deformed Loewner equation
are based on the observation that the Markov property is broken,
while scale invariance is not.
The definition for the upper half plane is
canonically extended to other simply connected domains $D$
by composition with a conformal map $\phi$
that sends $z=0$ and $z=\infty$ to two fixed boundary points of $D$
(canonical transport).
Therefore, conformal invariance in our case reduces to invariance
under the generic conformal automorphism
of $\mathbb{H}$ that keeps $z=0$ and $z=\infty$ fixed,
namely the scale transformation $z\mapsto \lambda z$ with $0<\lambda\in\mathbb{R}$.
The non-trivial property of the defining equation for $q$-SLE in the half plane is \emph{scale invariance};
the process is then coherently extended to other domains by conformal transport.
Consider the rescaled map
\beq\label{eq:rescaling}
\tilde g_t(z)=\lambda^{-1} g_{\lambda^2 t}(\lambda z),
\eeq
where $g_t(z)$ satisfies the $q$-SLE equation.
Computation of the time $q$-derivative of $\tilde g_t$ is straightforward thanks to
the fact that the usual chain rule is satisfied with respect to
linear homogeneous functions.
One has
\beq
\begin{aligned}
\qder{q}{t}\tilde g_t&=\lambda\, \qder{q}{\lambda^2 t} g_{\lambda^2 t}\\
&= \lambda \frac{2}{g_{\lambda^2 t}} - \lambda\, \qder{q}{\lambda^2 t}\xi_{\lambda^2 t}.
\end{aligned}
\eeq
By noting that, in law, $\xi_{\lambda^2 t}\sim \lambda\xi_t$, and again using the chain rule
for the $q$-derivative with respect to $\lambda^2 t$, one obtains
\beq
\qder{q}{t}\tilde g_t = \frac{2}{\tilde g_t} - \qder{q}{t}\xi_t.
\eeq
In order for $\tilde g$ to really satisfy the same equation as $g$, the initial condition
$\tilde g_{t\to 0}(z)=z$ must be met; it is straightforward to see that this is the case
(thanks to the $\lambda z$ in the definition of $\tilde g$).
Note that the scaling dimension $\frac{1}{2}$ of $g$, 
as can be read off from (\ref{eq:rescaling}),
is the same as for the classical SLE.

We now focus on the domain Markov property.
Let us first consider the ordinary SLE. 
Let $K_t$ be a family of growing hulls and
$g_t$ the translated Loewner map that sends the tip of the hull to the origin.
If we define $\hat g_{t,s} = g_{t+s}\circ g^{-1}_s$ as the map that absorbs
what remains of $K_{t+s}$ after the action of $g_s$, we see that formally
\beq\label{eq:markovproperty}
\partial_t \hat g_{t,s} = \left( \partial_{t+s} g_{t+s} \right)\circ g^{-1}_s = \frac{2}{\hat g_{t,s}} - \dot \xi_{t+s},
\eeq
which implies that $\hat g_{t,s}$ is equal in distribution to $g_t$ if $\xi_t$ is Brownian motion
(whose increments are invariant under time translations);
in other words, the process satisfies the domain Markov property.

The $q$-derivative, on the contrary, does not satisfy the chain rule with respect to
an inhomogeneous function, so that $\qder{q}{s+t}\neq\qder{q}{t}$.
In fact, for any function $h(\cdot)$, for small $s$,
\beq
\begin{aligned}
\qder{q}{t}h(t+s)&=\frac{h(qt+s)-h(t+s)}{(q-1)t}\\
&\approx \frac{h\left(q(t+s)\right) + (1-q)s\,h'\left(q(t+s)\right) - h(t+s)}{(q-1)t}\\
&=\left(1+\frac{s}{t}\right)\qder{q}{t+s}h(t+s)-\frac{s}{t}h'\left(q(t+s)\right)\\
&=\qder{q}{t+s}h(t+s)+\frac{s}{t}\left[\qder{q}{t+s}h(t+s)-h'\left(q(t+s)\right)\right],
\end{aligned}
\eeq
where we have expanded $h(qt+s)$ at first order around $q(t+s)$ in the second equality,
and used the definition of q-derivative in the third one.
If $g_{t+s}$ is a solution of (\ref{eq:qsle}), then the composed map $\hat g_{t,s}$ satisfies, for small $s$,
\beq
\begin{aligned}
\qder{q}{t}\hat g_{t,s}=&\frac{2}{g_{\sqrt{q}(t+s)}\circ g_s^{-1}}-\qder{q}{t+s}\xi_{t+s}\\
&+ \frac{s}{t}\left[ \qder{q}{t+s}g_{t+s}\circ g_s^{-1}-\partial_{q(t+s)} g_{q(t+s)}\circ g_s^{-1} \right],
\end{aligned}
\eeq
which, at first order in $s$, reads
\beq
\begin{aligned}
\qder{q}{t}\hat g_{t,s}=&\frac{2}{\hat g_{\sqrt{t},s}}-\qder{q}{t+s}\xi_{t+s}\\
&+\frac{s}{t}\left[ \frac{2}{\hat g_{\sqrt{q}t,s}}-\qder{q}{t+s}\xi_{t+s}-\partial_{qt}\hat g_{qt,s} 
+ 2t\left(1-\sqrt{q}\right)\frac{\partial_{\sqrt{q}t}\hat g_{\sqrt{q}t,s}}{\left(\hat g_{\sqrt{q}t,s}\right)^2}\right].
\end{aligned}
\eeq
The presence of the term in square brackets highlights the lack of Markov property.
It is clear from the computation that the Markov property would be broken
even if the right-hand side of equation (\ref{eq:qsle}) were evaluated at $t$ instead of $\sqrt{q}t$;
in this case the last term in the brackets would be absent.

Observe that the deformed Loewner equation, just like the classical one itself,
satisfies a simple reflection property.
In fact, writing (\ref{eq:qsle}) separately for the real and imaginary parts
of $g$ highlights the fact that if $g_t(z)$ is a solution with initial condition $z$,
then its conjugate $g^*_t(z)$ is a solution with initial condition $z^*$.
Therefore, the process is symmetric with respect to the real axis, which is
the justification for restricting to the upper half plane.
Another position would be to define the domain of growth as
the image of $\mathbb{H}$ under $g_t$.
Under this perspective, the domain would be a fluctuating
random non-fixed quantity,
which can be thought of as a manifestation of the ``quantum flavor'' of the
$q$-deformed LE.

\section{\label{section:special_cases}Special cases}

\subsection{\label{section:qslit}Constant driving function: the $q$-slit mapping}

In this section we are going to present the solution of the $q$-LE equation 
in the simplest case, \ie when the driving function is constant.
We recall that the solution to the classical LE in such case is the slit mapping
\beq\label{eq:classical_slit}
\phi_t(z)=\sqrt{z^2+4t},
\eeq
which we shall  rewrite temporarily as
\beq  
\label{eq:factorized_slit}
\phi_t(z)=z\sqrt{1+\frac{4t}{z^2}}.
\eeq
We seek to introduce a $q$-deformed version of this map; 
this can be obtained by considering the $q$-Pochhammer symbols:
\beq\label{eq:pochhammer}
(a;q)_k=\prod_{j=0}^{k-1}(1-a q^j),
\eeq
with $k>0$. We can then introduce the function
\beq\label{eq:psi_definition}
\psi^{(\alpha)}(x,q)=\frac{(-x;q)_\infty}{(-q^{\alpha}x;q)_\infty},
\eeq
which is regarded as a $q$-deformation of the function $(1+x)^\alpha$.
As such, it satisfies the following useful identities
(see propositions 14.1 and 14.2 in \cite{KacCheung:2002}):
\begin{eqnarray}
\label{eq:psi_derivative}\qder{q}{x} \psi^{(\alpha)}(x,q)=[\alpha]_q \psi^{(\alpha-1)}(qx,q),\\
\label{eq:psi_product}\psi^{(\alpha)}(x,q) \psi^{(\beta)}(q^{\alpha}x,q)= \psi^{(\alpha+\beta)}(x,q),
\end{eqnarray}
where $\left[\alpha\right]_q$ is defined as in (\ref{eq:qnumber}).
Relations (\ref{eq:psi_derivative}) and (\ref{eq:psi_product}) are direct generalizations of the analogue
identities for the undeformed case ($q=1$).

If we take $\alpha=\frac{1}{2}$ [so as to obtain the $q$-deformation of the square root in
(\ref{eq:factorized_slit})] we then have
\beq\label{eq:psi_deriv_1}
\qder{q}{x} \psi^{(\frac{1}{2})}(x,q)=\Big[\frac{1}{2}\Big]_q \psi^{(-\frac{1}{2})}(qx,q).
\eeq
By using (\ref{eq:psi_product}) and by rescaling $x$ one also has
\beq   \psi^{(\frac{1}{2})}(q^{\frac{1}{2}}x,q) \psi^{(-\frac{1}{2})}(qx,q)=1;
\eeq
therefore (\ref{eq:psi_deriv_1}) can be cast in the form
\beq  \qder{q}{x} \psi^{(\frac{1}{2})}(x,q)=\frac{\big[\frac{1}{2}\big]_q}{ \psi^{(\frac{1}{2})}(q^{\frac{1}{2}}x,q)}.
\eeq
If we now define the $q$-slit mapping as
\beq\label{eq:phi_definition}
\phi_t(z,q)= z \psi^{(\frac{1}{2})}\Big(\frac{2(1+\sqrt{q})t}{z^2},q\Big),
\eeq
we see that it satisfies the $q$-LE equation with no driving term
\beq  \qder{q}{t}   \phi_t(z,q)=\frac{2}{  \phi_{\sqrt{q}t}(z,q)},
\eeq
as can be readily checked by remembering that
\beq  \Big[\frac{1}{2}\Big]_q=\frac{1}{1+\sqrt{q}}.
\eeq

For later convenience we give the explicit expression of the map $\phi_t(z,q)$ in terms of infinite products,
from which one can read off at first sight the structure of zeroes and poles (see also Sec.~\ref{section:numerical_results}):
\beq\label{eq:phi_zeropoles}
\phi_t(z,q)= z \frac{\prod_{j=0}^{\infty}\Big(1+\frac{2(1+\sqrt{q})t}{z^2} q^j\Big)}{\prod_{j=0}^{\infty}\Big(1+\frac{2(1+\sqrt{q})t}{z^2} q^{j+\frac{1}{2}}\Big)}.
\eeq
We now want to study the expansion of this map around $z=\infty$. For this purpose we introduce the 
$q$-binomial coefficients:
\beq \binom{n}{m}_{\!q}= \prod_{i=0}^{m-1}\frac{1-q^{n-i}}{1-q^{i+1}}.
\eeq
By means of such quantities, the function $\psi^{(\alpha)}$ can be expanded in powers of $x$ as follows:
\beq\label{eq:psi_binomial}
\psi^{(\alpha)}(x,q)=\sum_{j=0} ^\infty \binom{\alpha}{j}_{\!q} q^{\frac{j(j-1)}{2}}x^j;
\eeq
this can be seen by expanding the definition (\ref{eq:psi_definition}) around $x=0$
(actually, the proof of this formula needs use of the so-called $q$-Taylor expansion;
for the details see \cite{KacCheung:2002}).
From (\ref{eq:psi_binomial}), it follows that the $q$-slit map has the following infinitesimal form
\beq  \phi_t(z,q)\sim z+\frac{2t}{z}+\frac{4\sqrt{q}t^2}{(1+q)z^3}
+O\left(\frac{t^3}{z^5}\right),
\eeq
which, at first order in $1/z$, is the same as the standard slit mapping; 
the higher orders, however, depend on $q$ as expected.
It can be checked that by taking the limit $q\to 1$ in the 
all-order expansion (\ref{eq:psi_binomial}) and by using the definition (\ref{eq:phi_definition})
one reproduces the well known expansion of the standard slit map.

Having obtained the analytical solution $\phi_t$, it is then straightforward to
determine the shape of the growing hull.
Keep in mind the definition (\ref{eq:hull}) of $K_t$, which is related
to the zeros of the map at time $\sqrt{q}t$.
From (\ref{eq:phi_zeropoles}) we find that the latter are given by
\beq z_{k}^2=-2(q+\sqrt{q})t q^{k}, \quad k=0,1,\ldots;
\eeq
they are thus all located along the imaginary axis, and they accumulate in a neighborhood of the origin. 
The zero which lies farthest from the origin is $z_0$, corresponding to $k=0$. 
If we fix time $t$, all points lying below $z_0$ on the imaginary axis have coincided with $z_0$
at a previous time $\tau\leq t$.
The hull $K_t$ thus coincides with a slit along the imaginary axis, just as in the classical case;
its length is
\beq \big|z_{0}\big|=\sqrt{2(q+\sqrt{q})t},
\eeq
which takes the expected value $2\sqrt{t}$ for $q=1$.

\subsection{\label{section:sqrt}Square-root driving function}

We study here the deformed Loewner equation with square root driving function, which reads
\beq\label{eq:squarerootdriving}
\qder{q}{t}g_t(z)=\frac{2}{g_{\sqrt{q}t}(z)}-\qder{q}{t}\sqrt{\kappa t},
\eeq
where $\kappa$ is a positive real constant.
This case is critical in the following sense.
If one considers a general driving function following a power law of exponent $\alpha$,
\beq
\xi_t = C t^\alpha,
\eeq
the same reasoning used in Sec.~\ref{section:properties}
shows that a rescaling of time such as in (\ref{eq:rescaling}) transforms the driving function into
\beq
\xi'_t = C \lambda^{2\alpha-1}t^\alpha.
\eeq
In other words, the multiplicative constant $C$ can always be rescaled away,
except in the critical case $\alpha=1/2$, which of course corresponds to the average behavior
of Brownian motion (see \cite{KagerNienhuisKadanoff:2004} for a discussion
of the classical case).

An estimation of the qualitative behavior of the solution can be obtained
by comparing the equation with its classical counterpart
for small times.
By performing the $q$-derivative of $\sqrt{t}$ in (\ref{eq:squarerootdriving}),
and by writing the other one explicitly, the equation becomes
\beq
\frac{g_{qt}(z)-g_t(z)}{(q-1)t}=\frac{2}{g_{\sqrt{q}t}(z)}-\frac{\sqrt{\kappa}}{1+\sqrt{q}}\frac{1}{\sqrt{t}}.
\eeq
The maps evaluated at times $qt$ and $\sqrt{q}t$ can be expanded (classically)
around $t$; 
by neglecting terms of order $t$
(while still keeping the singularity $1/\sqrt{t}$),
one obtains
\beq
\partial_t g_t(z)=\frac{2}{g_t(z)}-\partial_t \sqrt{\kappa_{\mathrm{eff}}t},
\eeq
which is the classical Loewner equation with square-root driving function with an effective
diffusion constant
\beq\label{eq:keff}
\kappa_{\mathrm{eff}}=\frac{4\kappa}{(1+\sqrt{q})^2}.
\eeq

It is useful to notice that when the driving function is exactly a square root
(as opposed to the average square root behavior of Brownian motion),
the scaling argument presented in Sec.~\ref{section:properties}
shows that the rescaled map is equal (as opposed to equal in law) to
the map itself:
\beq
g_t(z)=\lambda^{-1}g_{\lambda^2t}(\lambda z).
\eeq
A direct consequence of this is that the hull at all times
is the union of a (possibly infinite) number of straight lines
growing to infinity.

\section{\label{section:algorithm}Discretization and numerical algorithm}

In this section we propose a numerical scheme for the simulation of
the $q$-deformed Loewner equation 
(both the deterministic and stochastic versions).
The discrete nature of the $q$-derivative very naturally leads to a discretized
equation, which is equivalent to the ``continuous'' one; an approximation is
then introduced in order to deal with infinite series.

By applying the Jackson integral to both sides of (\ref{eq:qsle}), one obtains
\beq\label{eq:gjackson}
g_t(z)=2(1-q)t\sum_{j=0}^\infty \frac{q^j}{g_{q^{j+1/2}t}(z)} - \xi_t + z,
\eeq
where the integration constant $z$ accounts for the initial condition $g_{t\to 0}(z)=z$.
As explained in Sec.~\ref{section:definition}, the requirement of
continuity in $0$ fixes this constant uniquely.
We now introduce discretized times $0=t_0<t_1<\ldots<t_n=t$; in particular
we will fix $\Delta t$ and choose $t_i=i \Delta t$ in the following, in order to simplify the equations.
\begin{figure}[t]
\centering
\includegraphics[scale=0.8]{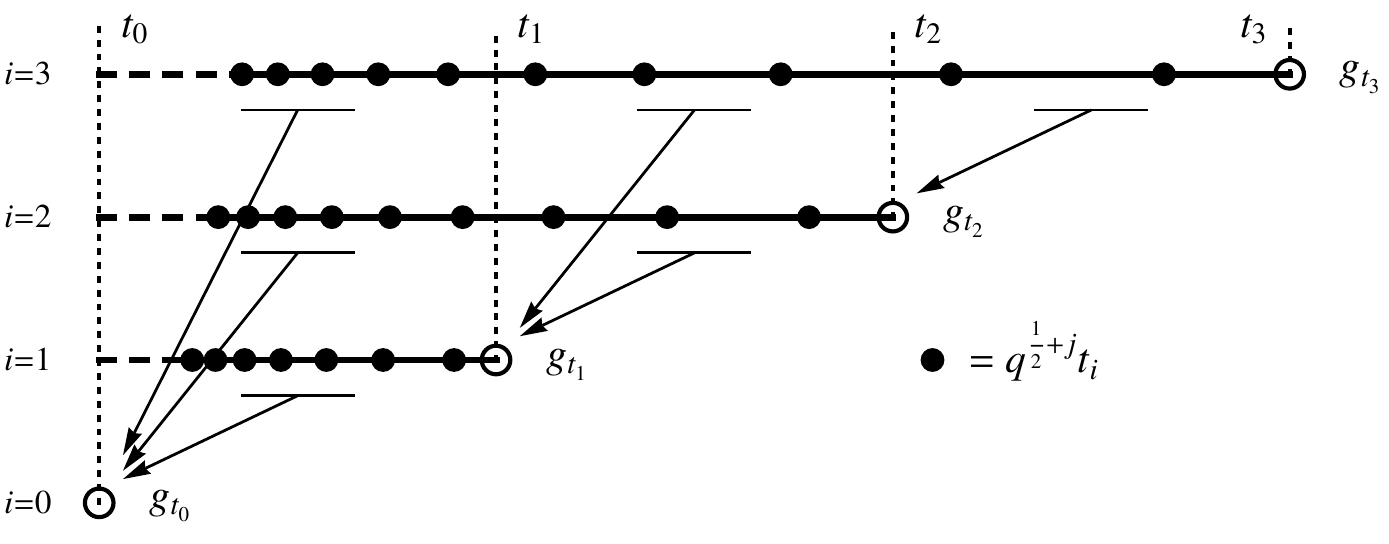}
\caption{
Schematic description of the algorithm.
The initial condition is specified at step $i=0$ (bottom left corner).
At each increasing time step (vertical axis), the computation of the current variable
(empty circles) would need the variables (filled disks) at an infinite number of past times
(horizontal axis), which are approximated by their closest known values (arrows).
}
\label{figure:algorithm}
\end{figure}
Let us focus on the $i$-th time~$t_i$. The sum over $j$ contains the variables $g$
evaluated at previous times $q^{j+1/2}t_i$; the first of these values is strictly
less than $t_i$ and they accumulate towards $t_0=0$ (remind that $q<1$).
The approximation we introduce is the following: the variables at these intermediate times
are taken to be equal to the closest ones (in time) that have been already computed at
previous steps $t_0,\ldots,t_{i-1}$.
Refer to Fig.~\ref{figure:algorithm}.
At the first step $i=0$ we have the initial condition $g_{t_0}(z)=z$.
At $i=1$, none of the variables appearing in the sum have been computed,
so we take them all equal to $g_{t_0}(z)=z$;
the fact that this is a sensible approximation relies on the requirement of
continuity for $g_t$ in $t=0$.
Thus we have
\beq
\begin{aligned}
g_{t_1}(z)&= 2(1-q)t_1\sum_{j=0}^\infty \frac{q^j}{g_{t_0}(z)}-\xi_{t_1}+z\\
&=z+\frac{2}{z}t_1-\xi_{t_1},
\end{aligned}
\eeq
where the geometric sum has been performed in the second equality.
At $i=2$, the variables at a finite number of intermediate times will be approximated
by $g_{t_1}$ --- say those from $j=0$ to some maximal $j$ in the sum --- while
an infinite number thereof will still have to be set equal to $g_{t_0}$.
In general, at the $i$-th step, a finite number of variables will be approximated as
$g_{t_{i-1}}$, a finite number as $g_{t_{i-2}}$, and so on; the number of variables
approximated as $g_{t_0}$ is always infinite.
Let $\sigma_k(i)$ (for $0<k<i$) be the smallest index such that
\beq
q^{\sigma_k(i)+1/2}t_i\leq t_k;
\eeq
it can be readily computed as
\beq
\sigma_k(i)=\Bigg\lfloor \log_q\left(1-\frac{k}{i}\right)-\frac{1}{2}\Bigg\rfloor,
\eeq
where $\lfloor\cdot\rfloor$ denotes the integer part.
Remark that the formula implies $\sigma_0(i)=-1$ and
it can be extended also to $\sigma_i(i)=\infty$,
by taking the limit $k\to i$. 
The $i$-th variable can then be written as
\beq
\begin{aligned}
g_{t_i}(z)=2(1-q)t_i&\left[
\sum_{j=0}^{\sigma_1(i)}\frac{q^j}{g_{t_{i-1}}(z)} +
\sum_{j=\sigma_1(i)+1}^{\sigma_2(i)} \frac{q^j}{g_{t_{i-2}}(z)} \right. \\
&\left.+\cdots +
\sum_{j=\sigma_{i-1}(i)+1}^{\infty} \frac{q^j}{g_0(z)}
\right] - \xi_{t_i} + z.
\end{aligned}
\eeq
Sums of $q^j$ can now be performed; a compact form of the result is
\beq\label{eq:algorithm}
g_{t_i}(z)=2 t_i \sum_{k=1}^i \frac{1}{g_{t_{i-k}}(z)} 
\left[ q^{\sigma_{k-1}(i)+1}-q^{\sigma_k(i)+1}\right] -\xi_{t_i}+z,
\eeq
where the boundary values of $\sigma$ are those described above, and $q^{\infty}=0$.

It is interesting to compare this algorithm with the
corresponding one for the classical case.
Let $\tilde g_t$ be the solution to
\beq
\rmd \tilde g_t(z)=\frac{2}{\tilde g_t(z)}\rmd t-\rmd \xi_t
\eeq
with initial condition $g_0(z)=z$ (and $\xi_0=0$); integrating from $0$ to $t$ gives
\beq
\tilde g_t(z) - z = \int_0^t \frac{2}{\tilde g_\tau(z)} \rmd\tau - \xi_t.
\eeq
By introducing discrete times $t_k$ as before and by
replacing the integral with a sum one obtains the following approximation:
\beq\label{eq:algorithmq1}
\begin{aligned}
\tilde g_{t_i}(z)&= 2 \sum_{k=0}^{i-1} \frac{1}{\tilde g_{t_k}(z)}-\xi_{t_i}+z\\
&=2 \sum_{k=1}^i \frac{1}{\tilde g_{t_{i-k}}(z)} -\xi_{t_i}+z.
\end{aligned}
\eeq
This equation differs from (\ref{eq:algorithm}) only by the term 
$q^{\sigma_{k-1}(i)+1}-q^{\sigma_k(i)+1}$, and by the overall factor $t_i=i\Delta t$.
By writing $\lfloor x\rfloor=x-\{x\}$ in the definition of $\sigma_k(i)$, where
$\{x\}\in[0,1)$ is the fractional part of $x$, one readily verifies that
\beq\label{eq:qlimit1}
\lim_{q\to 1}q^{\lfloor \log_q\left(1-\frac{k-1}{i}\right)-\frac{1}{2}\rfloor}-
q^{\lfloor \log_q\left(1-\frac{k}{i}\right)-\frac{1}{2}\rfloor}=\frac{1}{i},
\eeq
so that (\ref{eq:algorithm}) reduces to (\ref{eq:algorithmq1}).
The function in the left-hand side of (\ref{eq:qlimit1}) at fixed $q$ and $i$ is $0$ everywhere,
apart from some spikes that accumulate towards $k=i$, 
whose values decrease with increasing $k$, 
and whose number increases when $q$ approaches $1$.
This structure is reminiscent of what we see in numerical
simulations (see Sec.~\ref{section:numerical_results}),
and is descriptive of the complicated way that the $q$-deformed equation
converges to the classical one.

The Loewner equation is usually simulated with a different strategy \cite{Kennedy:2009review}.
It can be seen that the inverse map $\tilde g_t^{-1}(z)$, though
satisfying a partial differential equation involving its derivative
with respect to $z$ as well as its time derivative, 
is the same, in law, as $\tilde g_{-t}(z)$ at fixed $t$,
modulo a time-reversal of the driving function, which is
irrelevant for Brownian motion
(see also \cite{ChenRohde:2009}).
This is very convenient, because one can then easily obtain the tip
of the curve, which is defined as $\tilde g^{-1}_t(0)$.
In the deformed case, however, such a simple relation does not hold,
as a consequence of the broken Markov property.
Moreover, we are interested in looking at global properties of the map,
therefore the strategy we employ for simulating the hulls
is less direct but more general.
We discretize a rectangular portion of the half plane with a rectangular lattice,
and explicitly iterate equation (\ref{eq:algorithm}) for every
single point $z$ in the lattice.
Since we are interested in those points that got mapped to $z=0$
at some instant before the present time,
we keep track of the closest distance to the origin that the evolution
has reached up to time $t$:
\beq
\label{eq:minimumdistance}
d_t(z)= \inf_{\tau\leq t} \left|g_\tau(z)\right|.
\eeq
Plotting this quantity against the two-dimensional value of $z$ then highlights
the qualitative features of the hull (see Sec.~\ref{section:numerical_results}), 
which is defined as the set of all points for which $d_t(z)=0$.

It should be straightforward to apply the algorithm presented in this section
to other kinds of first-order $q$-deformed differential equations.
Actually, it should be possible to extend it also to partial-differential equations
with minimal effort, as long as the derivatives in the ``space'' dimensions
are ordinary.
Notice that the strategy of calculating the numerical evolution of all points
in a lattice is very straightforwardly vectorized for efficient parallel computing.
The simulations we present in the next section have been written
in the CUDA C/C++ programming language \cite{CUDAbyExample}
and run on the GPU NVIDIA GeForce GTX 480
(\texttt{http://www.nvidia.com/object/cuda\_home\_new.html}).
Computation time is of order one hour
for a lattice of $1024\times 1024$ points and $5000$ iterations.

\section{\label{section:numerical_results}Numerical results}
\begin{figure}[t]
\centering
\subfigure[$q=1$]{
\includegraphics[scale=0.9]{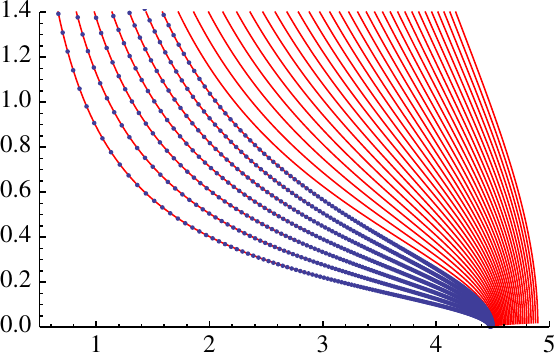}
}
\hspace{0.5cm}
\subfigure[$q=0.9$]{
\includegraphics[scale=0.9]{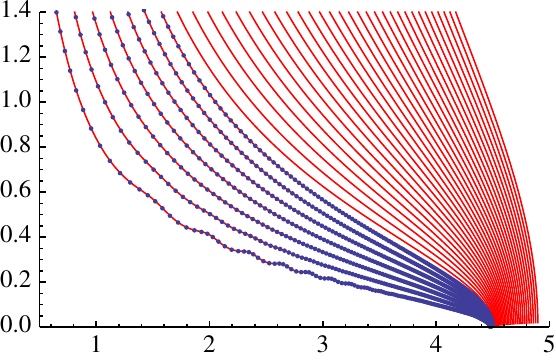}
}
\subfigure[$q=0.8$]{
\includegraphics[scale=0.9]{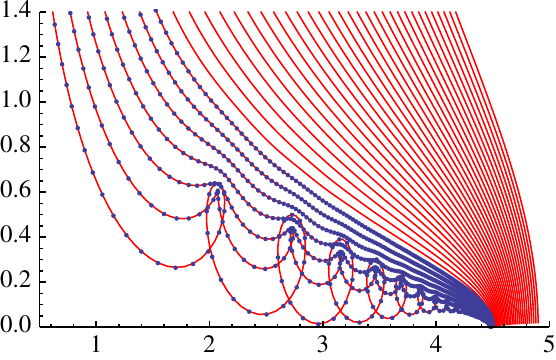}
}
\hspace{0.5cm}
\subfigure[$q=0.7$]{
\includegraphics[scale=0.9]{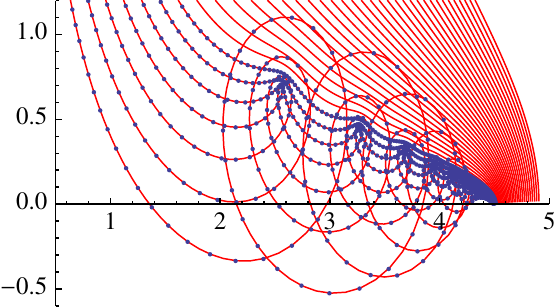}
}
\caption{
(Color online)
Images of vertical equally-spaced lines,
with real parts ranging from $0.2$ to $2$,
under the $q$-slit mapping at $t=5$.
Red lines are the analytical solution (\ref{eq:phi_zeropoles}),
blue points are the results of simulations (5000 iterations).
Notice that the agreement does not deteriorate close to the singularities.
The plot for $q=1$ (which can not be simulated directly with the algorithm presented,
because of divergences)
has actually been obtained by setting $q=0.99999$; it is checked against
the theoretical curves for the classical slit (\ref{eq:classical_slit}).
}
\label{figure:levellines}
\end{figure}

In order to check the validity and precision of the algorithm 
proposed in the previous section
we compare the exact solution for constant driving function
with the map obtained numerically as in (\ref{eq:algorithm}).
Figure \ref{figure:levellines} shows the images,
under the map $g_t$ at a fixed time $t$, of vertical equally-spaced
lines in the first quadrant.
The algorithm (after $5000$ iterations) yields a very accurate
representation of the map.
As is clear from the figure, the $q$-slit mapping is not
1-to-1, and there are points whose image has negative imaginary part.
In fact, the form (\ref{eq:phi_zeropoles}) reveals that an infinite
number of poles (and of zeroes) lie on the real axis;
moreover, they accumulate towards the essential singularity $z=0$.
This is in contrast with what happens in the classical Loewner
theory, where the maps are conformal, but it does not pose problems
for the definition (\ref{eq:hull}) of growing hull.

Let us now turn to the numerical solution of the equation with
square root driving function.
\begin{figure}[t]
\centering
\subfigure[$\kappa=2$, $q=0.7$]{
\includegraphics[scale=0.43]{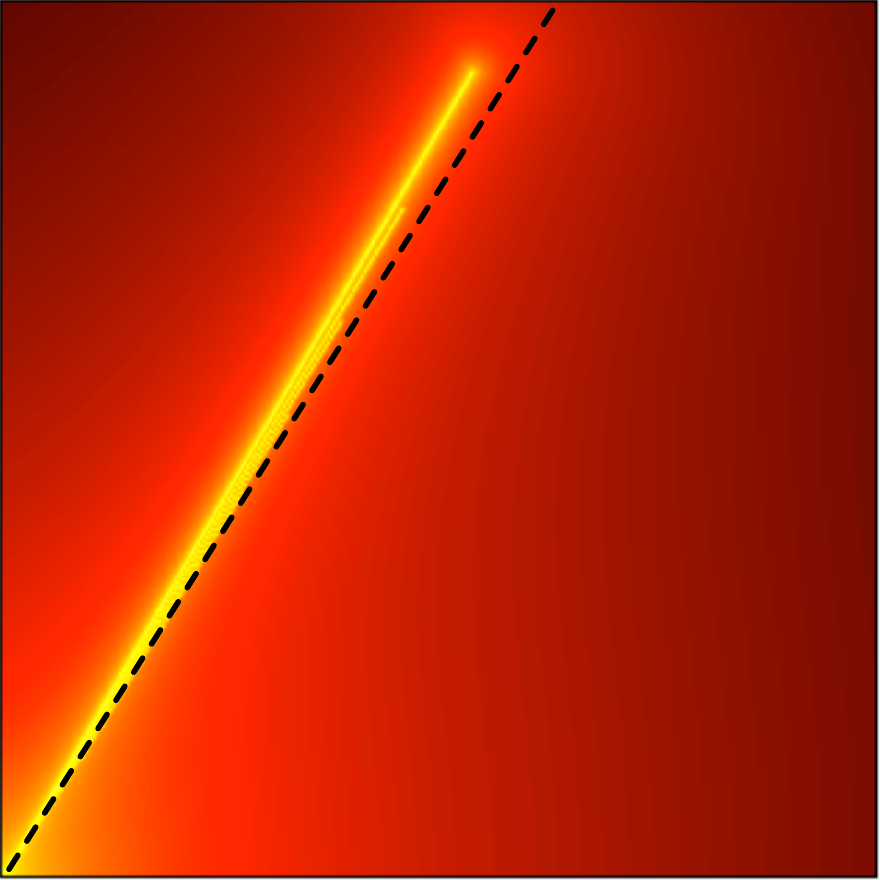}
}
\subfigure[$\kappa=2$, $q=0.5$]{
\includegraphics[scale=0.43]{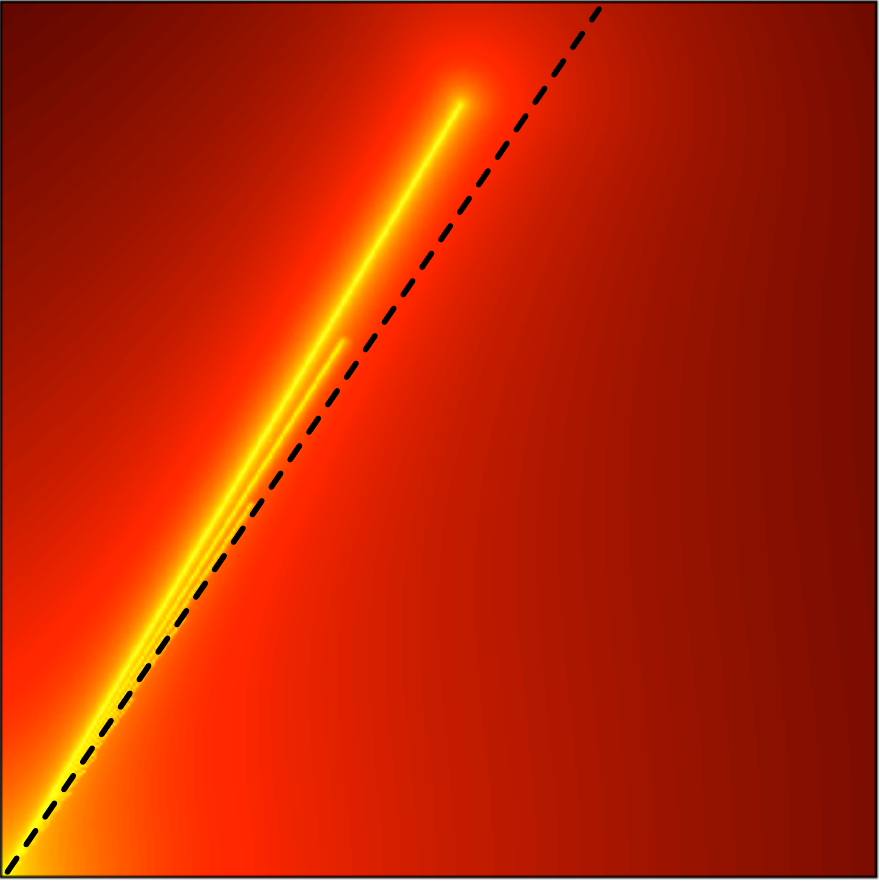}
}
\subfigure[$\kappa=2$, $q=0.3$]{
\includegraphics[scale=0.43]{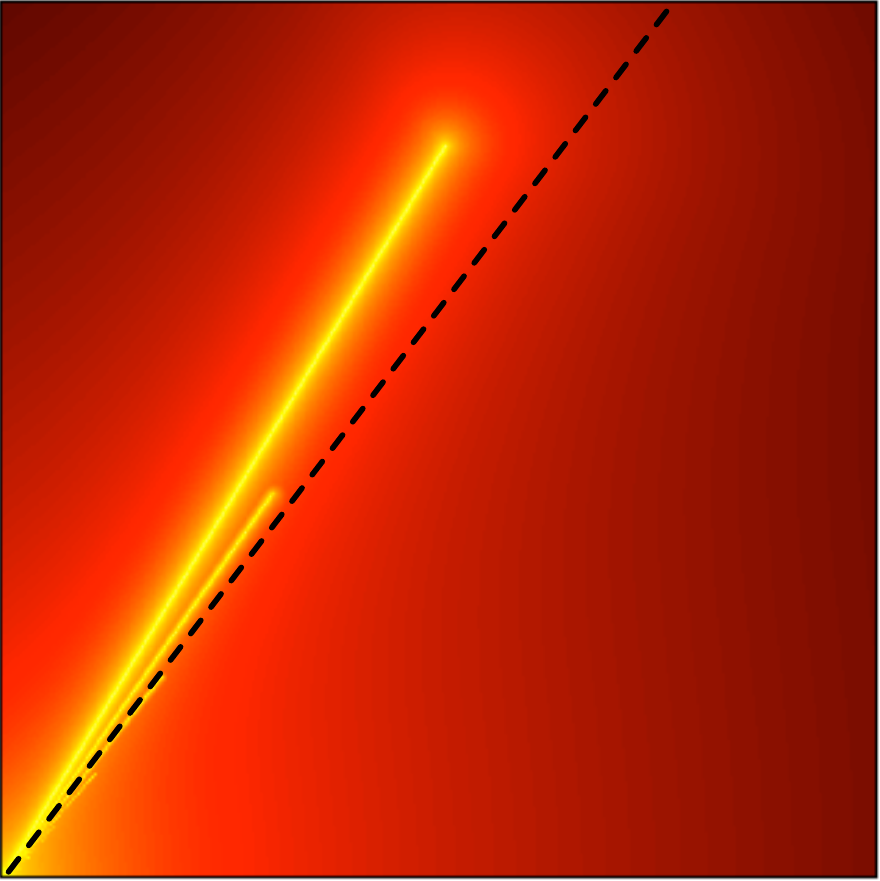}
}
\subfigure[$\kappa=6$, $q=0.7$]{
\includegraphics[scale=0.43]{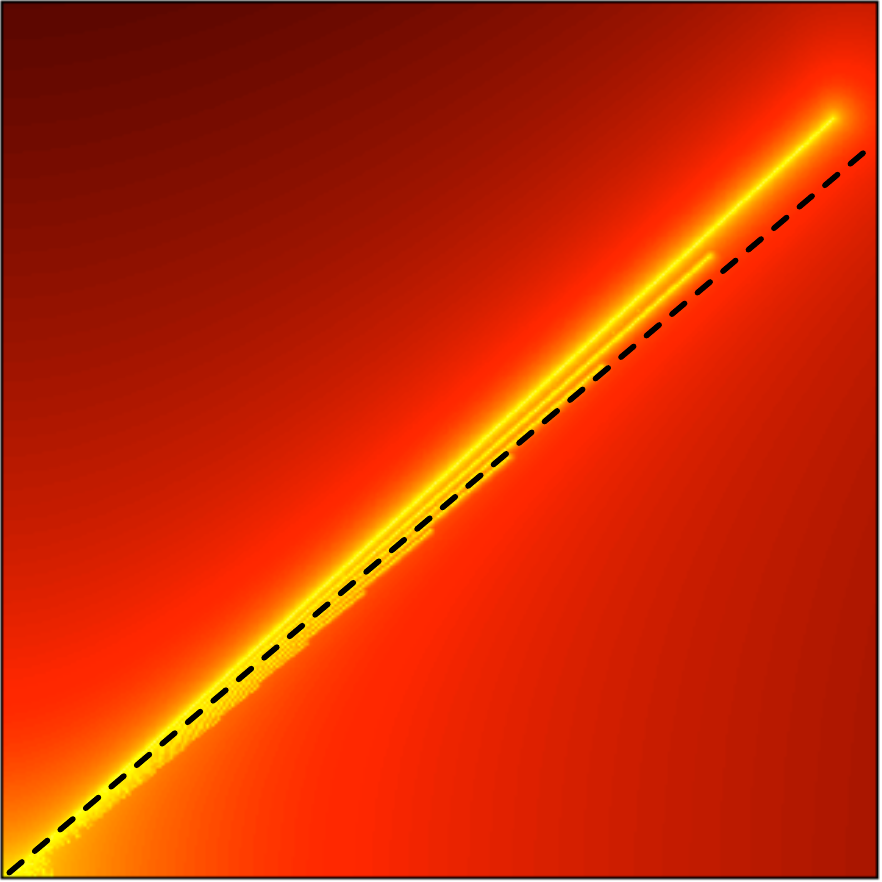}
}
\subfigure[$\kappa=6$, $q=0.5$]{
\includegraphics[scale=0.43]{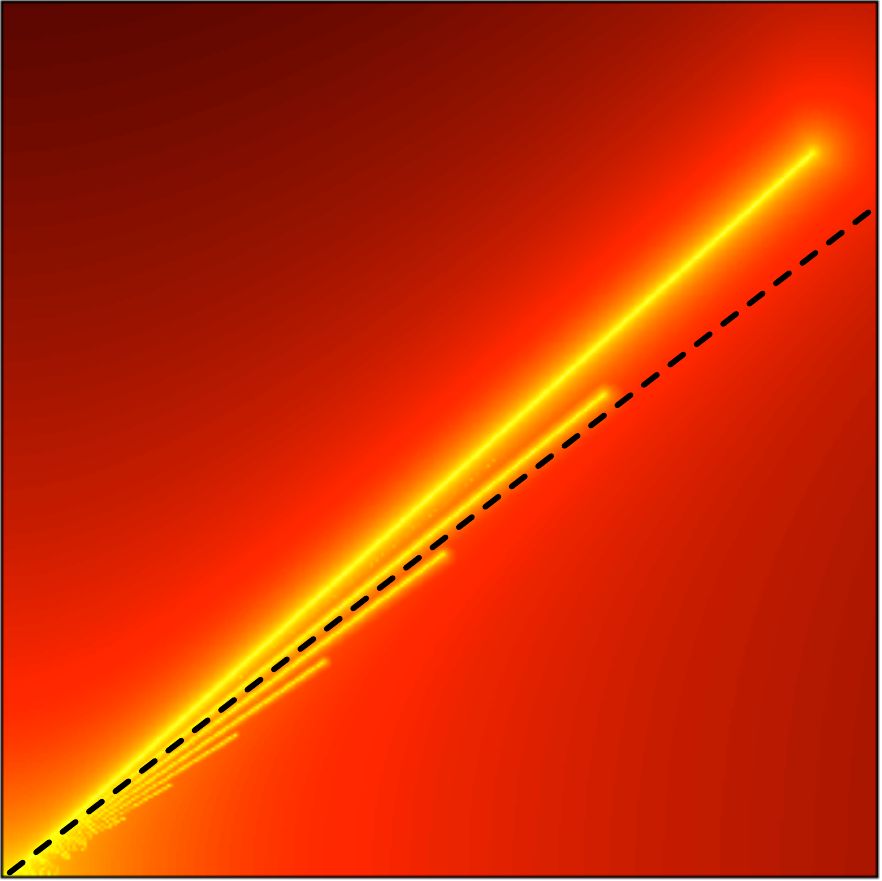}
}
\subfigure[$\kappa=6$, $q=0.3$]{
\includegraphics[scale=0.43]{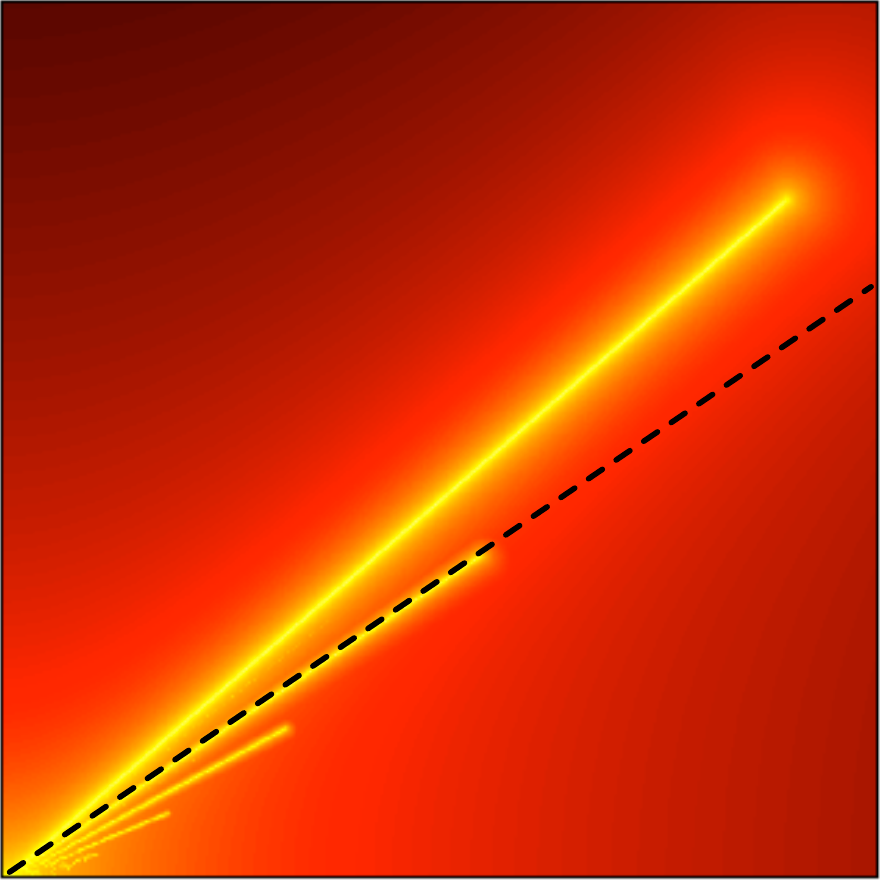}
}
\caption{(Color online)
The square root driving function gives rise to multiple lines.
The plots show the minimum distance $d_t(z)$ that the evolution starting from
$z$ has come to the origin up to a fixed time $t=10$, for $z$ in the upper-right quadrant
(positive real and imaginary part). Brighter colors correspond to
smaller values; the range is $0\apprge d_t(z)\apprge 7$.
Dashed lines show the approximate behavior given by (\ref{eq:effectivetheta}),
corresponding to the classical equation with a rescaled driving function.
}
\label{figure:squareroot}
\end{figure}
The corresponding classical solution is well known
\cite{KagerNienhuisKadanoff:2004,Kennedy:2009review}
and generates a growing segment forming an angle $\theta(\kappa)$
with the horizontal.
From the discussion in Sec.~\ref{section:sqrt}, we expect to find a (possibly infinite)
number of segments, whose ``average'' behavior, in some sense, should
be similar to the classical Loewner map with driving function
$(\kappa_\mathrm{eff} t)^{1/2}$, with $\kappa_\mathrm{eff}$ given by (\ref{eq:keff}).
The angle $\theta(\kappa)$ can be calculated exactly for $q=1$:
\beq
\theta(\kappa)= \frac{\pi}{2}\left(1-\sqrt{\frac{\kappa}{\kappa+4}}\right).
\eeq
In the $q\neq 1$ case we therefore define an effective angle 
$\theta_\mathrm{eff}(\kappa,q)=\theta(\kappa_\mathrm{eff})$:
\beq\label{eq:effectivetheta}
\theta_\mathrm{eff}(\kappa,q)=\frac{\pi}{2}\left(1-\frac{1}{2}\sqrt{
\frac{\kappa}{1+\kappa/4+2\sqrt{q}-q}}\right).
\eeq
\begin{figure}[t]
\centering
\includegraphics[scale=0.4]{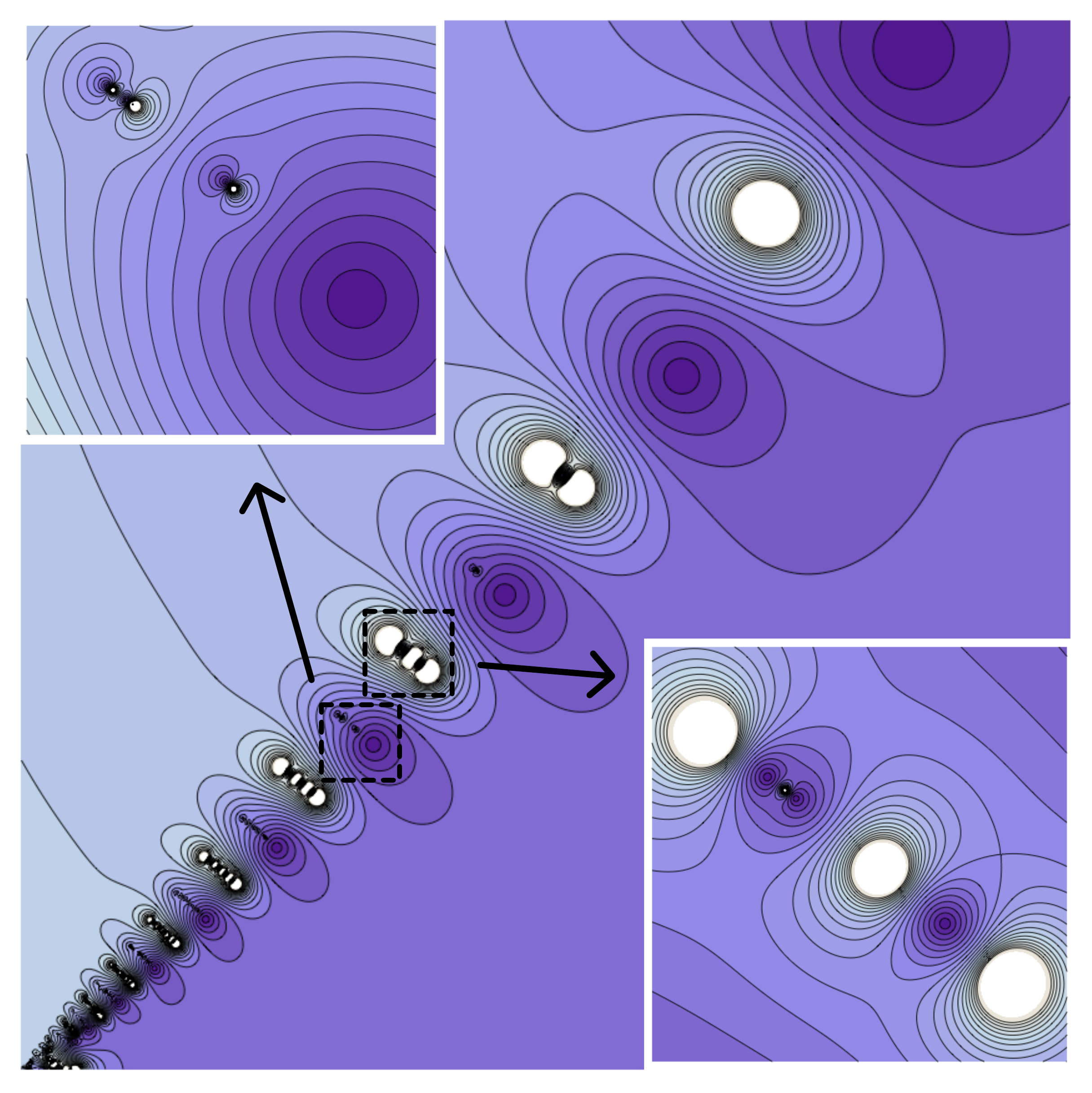}
\caption{
(Color online)
Contour plot of the numerical solution to the $q$-deformed Loewner equation
with square-root driving function, showing a rich structure
of zeros and singular points.
Darker colors correspond to complex numbers with smaller moduli.
The two insets zoom into two lines of zeros and poles.
}
\label{figure:poleszeros}
\end{figure}
Figure~\ref{figure:squareroot} gathers plots of the function $d_t(z)$
defined in (\ref{eq:minimumdistance}) for a fixed diffusivity $\kappa$
and several values of $q$, together with the corresponding lines 
of slope given by (\ref{eq:effectivetheta}).
We witness the emergence of multiple line segments starting
from the origin and growing in different directions, the lower ones
being the shorter.
Their slope gets steeper as $\kappa$ decreases towards $0$,
where they have to approach the $q$-slit solution.
For a fixed $\kappa$, increasing $q$ makes the branches
closer to each other, while their lengths become less different.
The picture that we get is that for $q\to 1$ all the segments reach
the same length, and they superimpose on each other, thus
reproducing the single-segment $q=1$ solution.

\begin{figure}[t]
\centering
\subfigure{
\includegraphics[scale=0.61]{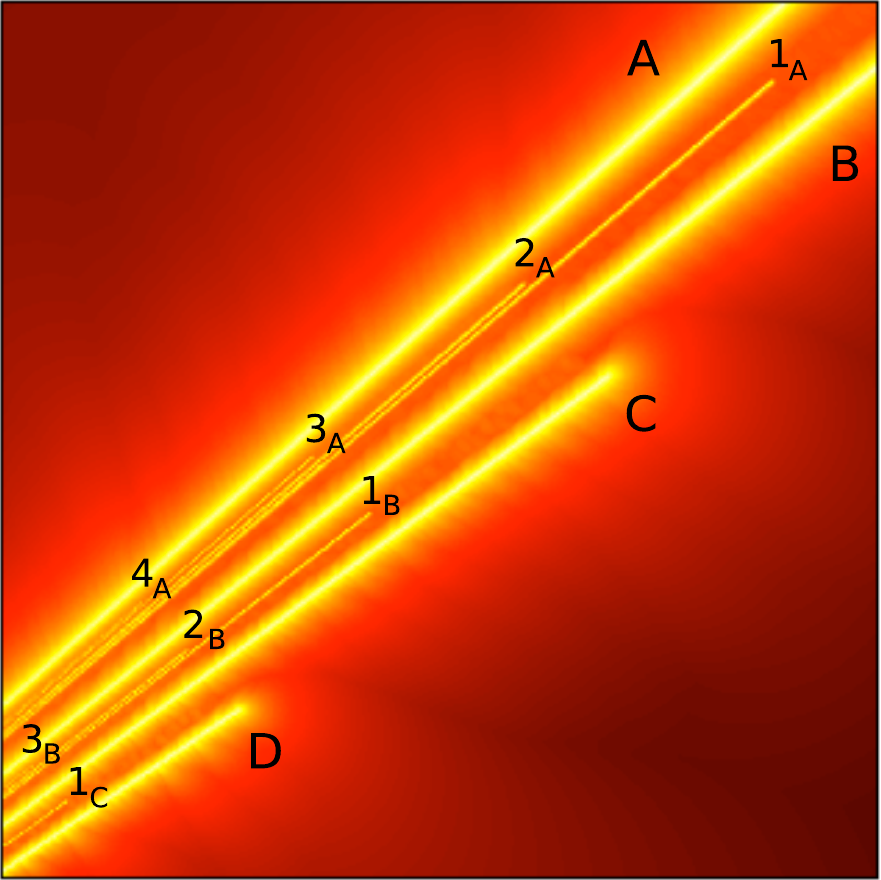}
}
\subfigure{
\includegraphics[scale=0.50]{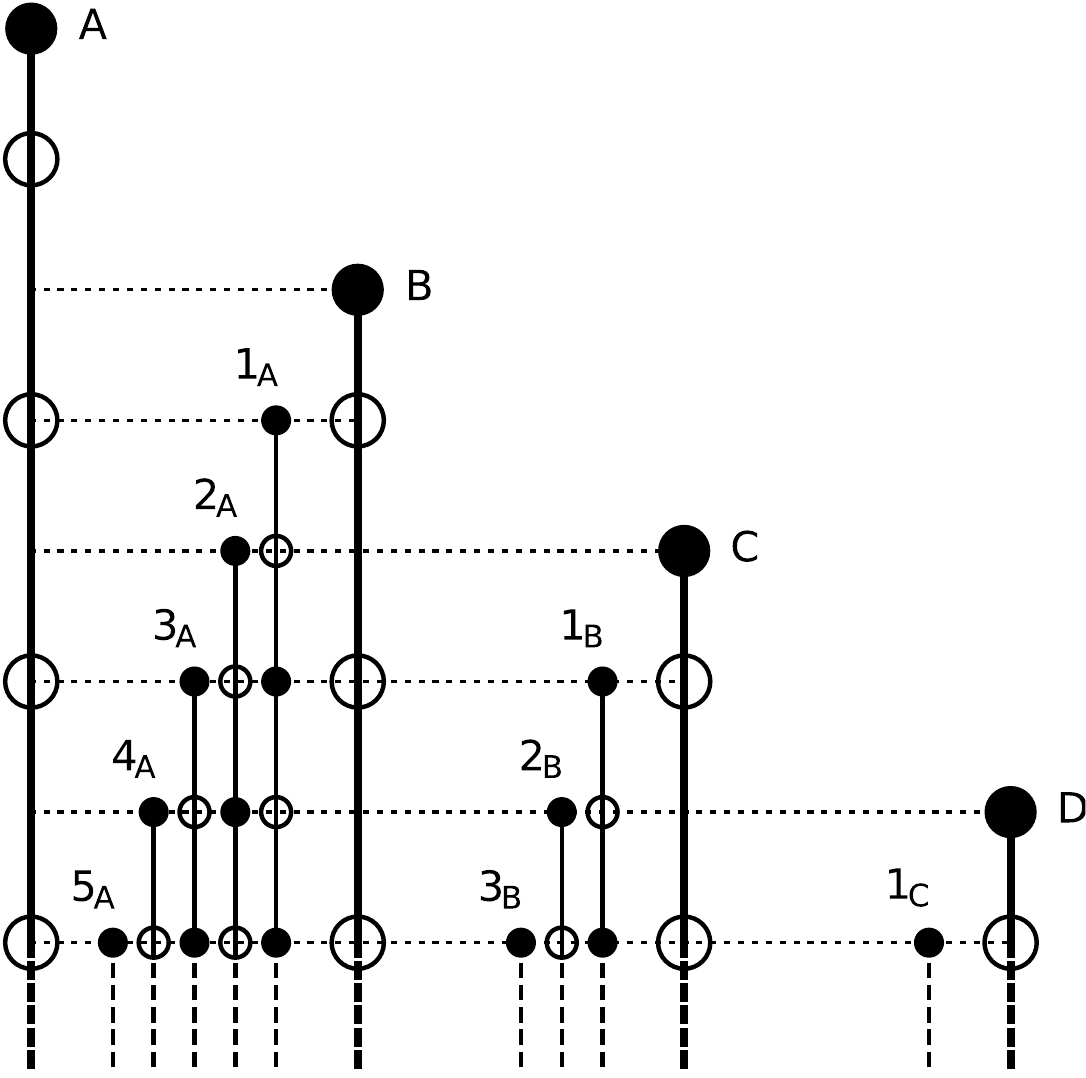}
}
\caption{
(Color online)
The multiple slits arising from the square-root driving function
(left), with $q=0.5$ and $\kappa=6$,
and the inferred structure of poles (empty circles) and zeros (filled disks)
of the map (right).
The first four main branches (denoted A, B, C, D) and a few of their
respective secondary branches ($1_\mathrm{A}$, $2_\mathrm{A}$, etc.)  
are labeled in both plots.
The plot on the left shows $\delta_t(z)$, defined in (\ref{eq:maxder}), at $t=10$;
the range is $0\apprge \delta_t(z)\apprge 10^6$.
Brighter colors correspond to higher values.
}
\label{figure:subbranches}
\end{figure}

Plots of $d_t(z)$ are handy but they can hide parts of the hull.
This is due to the smearing of the areas for which $d_t(z)\approx 0$.
Consider for instance a point $z$ lying deep inside a fjord between 
the branches in Fig.~\ref{figure:squareroot}; the evolution brings it
very close to the origin owing to the fact that $z$ is very close to
one of the visible branches.
But $z$ itself could also belong to another hidden branch, whose
$d_t(z)$ converges to zero less quickly as the number of
discretization steps $n$ grows.
In fact, if one looks directly at a plot of $\left|g_t(z)\right|$, which is
the distance from the origin that the point has come at time $t$,
a much richer structure is revealed; see Fig.~\ref{figure:poleszeros}.
Poles and zeros of $g_t$ lie, as expected, on straight lines, and accumulate
towards the origin.
A numerical analysis shows that the main branches that are visible 
in Fig.~\ref{figure:squareroot} correspond to lines that start with a zero,
say at point $z_j$, and then present poles at the points
$q^{k+1/2}z_k$ for $k=0,1,\ldots$.
But in between these branches, more and more zero-pole couples
appear as one approaches the origin. These zeros are not collinear,
so they give rise to different branches belonging to the hull.
These secondary lines (which are not visible in the plots of $d_t(z)$),
can be made to emerge by considering the modified observable
\beq\label{eq:maxder}
\delta_t(z)= \sup_{\tau\leq t} \left|\frac{\rmd}{\rmd\tau} g_\tau(z)\right|
\approx \frac{1}{\Delta t}\max_{i\leq n} \left| g_{t_i}(z)-g_{t_{i-1}}(z) \right|.
\eeq
Figure~\ref{figure:subbranches} shows a plot of this quantity in a region
close to the first four main branches; in between the latter, smaller branches
appear: finer simulations (not shown) expose
an infinite number of sub-branches in each fjord, whose lengths
decrease as their directions approach that of the main branch
(the fjord's coast line).
A schematic structure for zeros and poles inferred from the simulations
is also shown in Fig.~\ref{figure:subbranches}.
We stress that the results described in this section are numerical,
and we do not have mathematical proofs of the statements we make.
The main branches are composed of one zero and an infinite number
of poles, whereas the secondary branches are made of
an infinite number of alternating zeros and poles.
All these points lie on lines which are approximately perpendicular to
the leftmost main branch, in such a way that poles and zeros alternate
on such lines as well, also including those lying on the main branches.

The map for the square-root driving function is expected to
converge to the one for the constant driving function (\ref{eq:phi_zeropoles})
as $\kappa$ goes to zero.
This explains why secondary poles and zeros appear
in couples: as $\kappa$ decreases, every line converges towards
the imaginary axis, and each of these couples annihilates.
This behavior can be reproduced by collapsing the horizontal lines in the schematic
representation of Fig.~\ref{figure:subbranches}, \ie by superimposing
all vertical lines.
What is left of each line is one zero (its tip) and one pole
(just below its tip), which gives rise to the alternating zero-pole
pattern of (\ref{eq:phi_zeropoles}).
Conversely, an interesting (and more difficult) question is how this picture converges to
the classical one as $q$ goes to $1$, for fixed $\kappa$.
The solution to the classical Loewner equation with square-root driving function
at fixed time has a single zero at the tip of the line.
As one increases $q$ towards $1$ (see Fig.~\ref{figure:squareroot})
all the branches come closer to each other, and their lengths
converge towards a common value (although not uniformly; namely, for any $q$
and any $\epsilon$ there is always an infinite number of branches of
length less than $\epsilon$).
On each branch, the density of singular points increases, and zero-pole
couples on the secondary branches get closer.
What happens is that each pole will be compensated by exactly one zero,
and only zeros (possibly an infinite number thereof) remain in the end,
all converging to the tip of the classical single-slit solution.

\begin{figure}[t]
\centering
\subfigure[$q=0.9$]{
\includegraphics[scale=0.35]{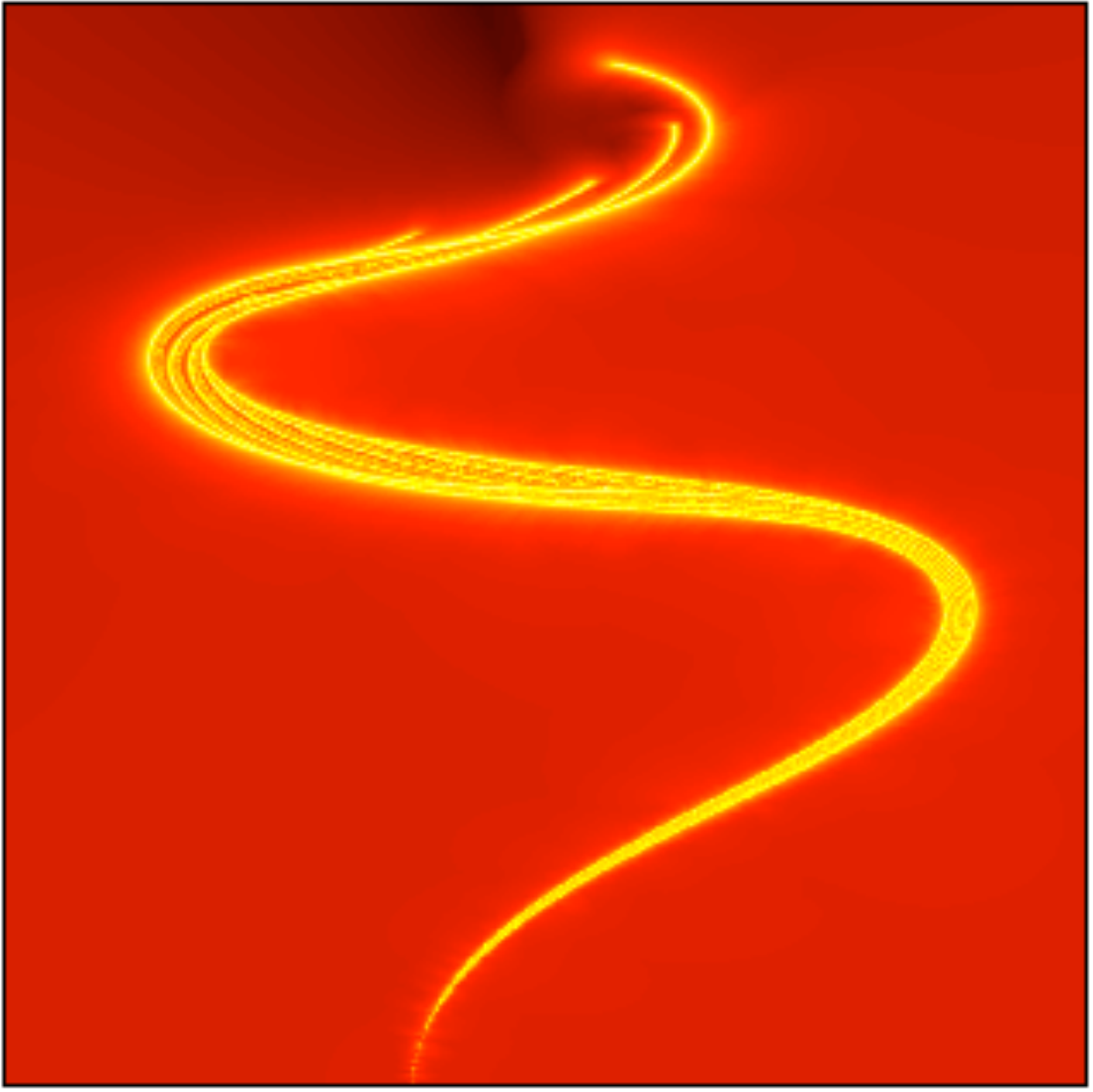}
}
\subfigure[$q=0.7$]{
\includegraphics[scale=0.35]{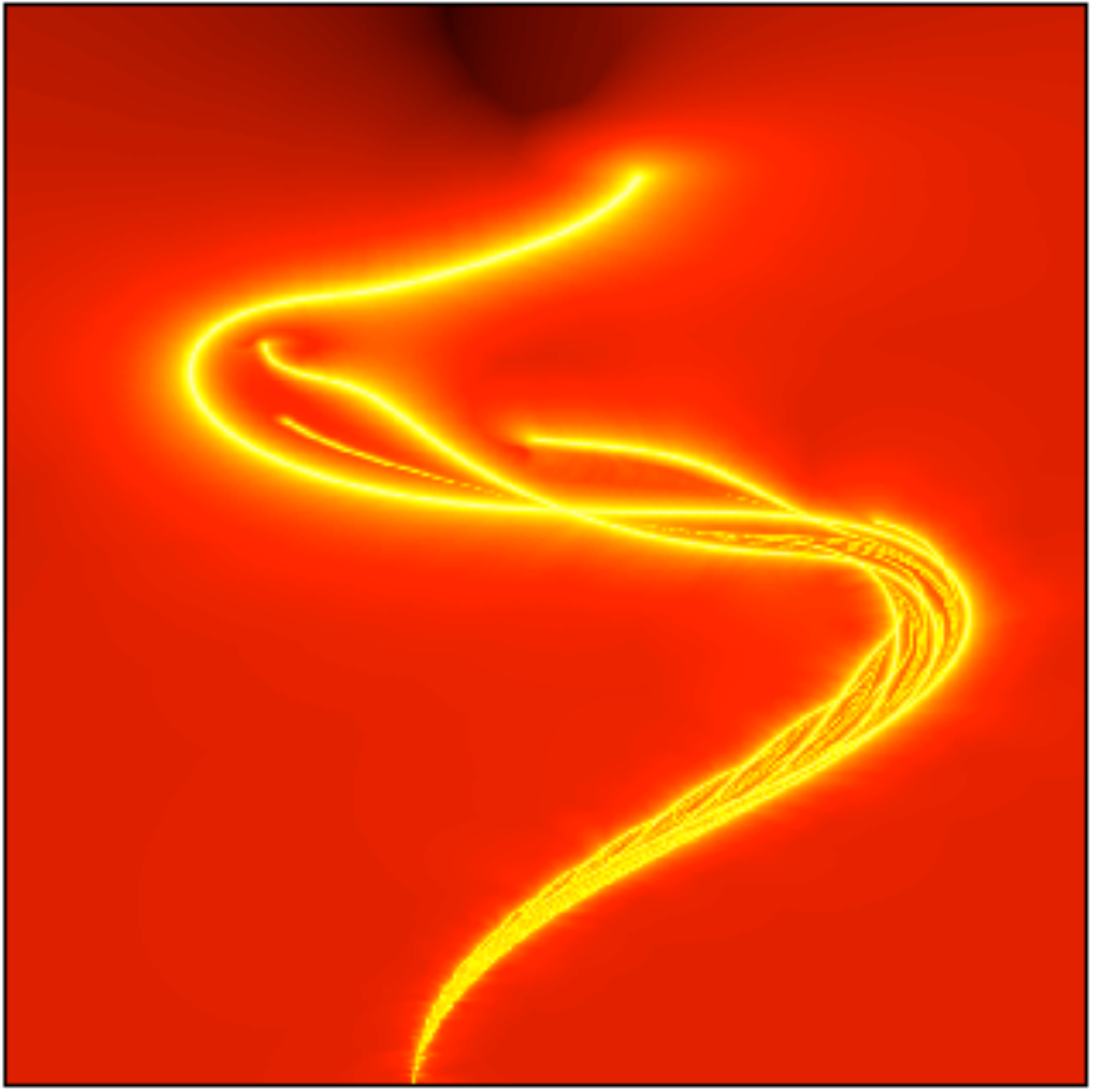}
}
\subfigure[$q=0.5$]{
\includegraphics[scale=0.35]{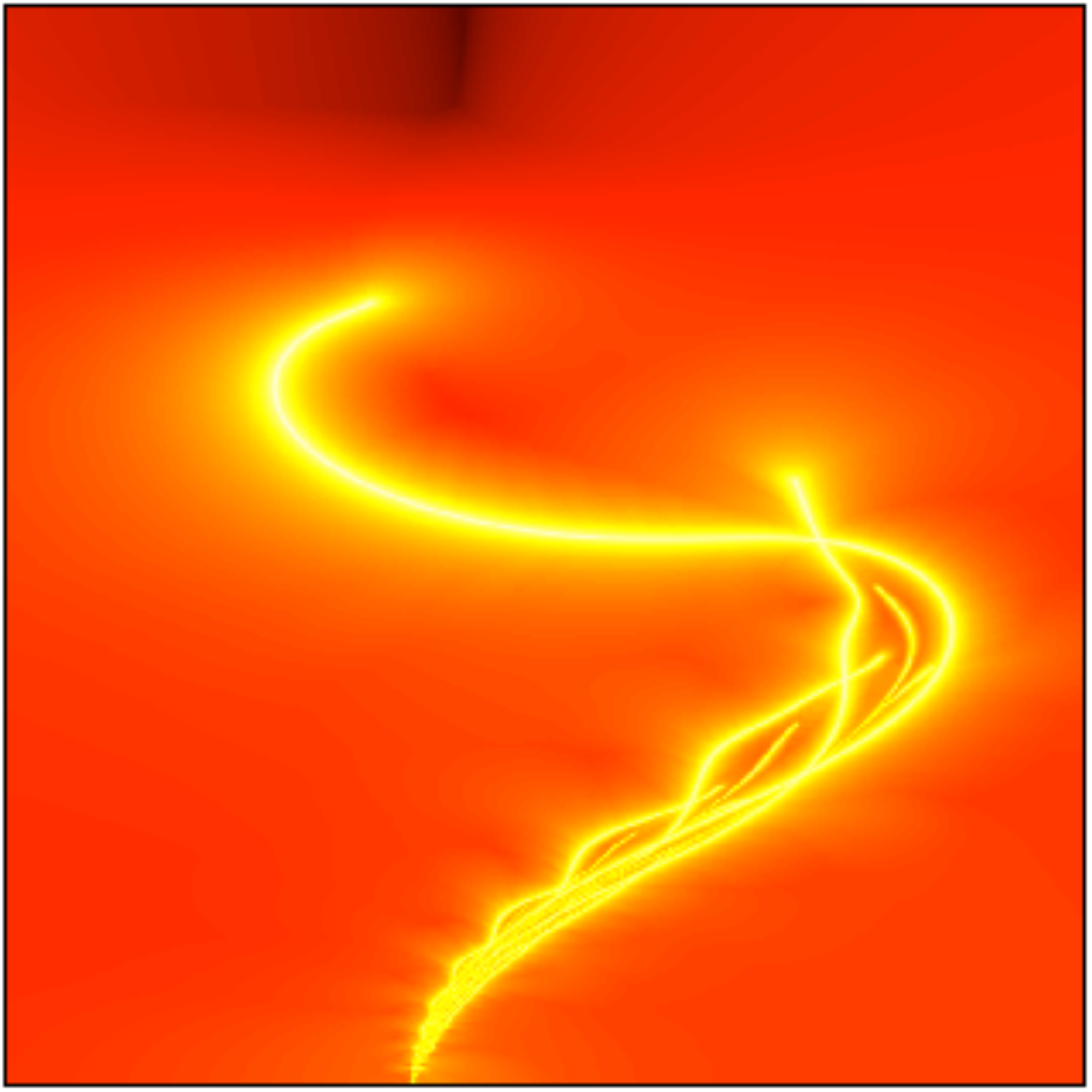}
}
\caption{
(Color online)
A non-monotonic driving function (sinusoidal in this case)
produces intersecting curves, which get further apart
when decreasing $q$.
The plot shows $\delta_t(z)$, defined in (\ref{eq:maxder}), for a fixed time $t$, 
with brighter colors corresponding to higher values.
}
\label{figure:sin}
\end{figure}

\begin{figure}[ht]
\centering
\subfigure[$q=0.99$]{
\includegraphics[scale=0.35]{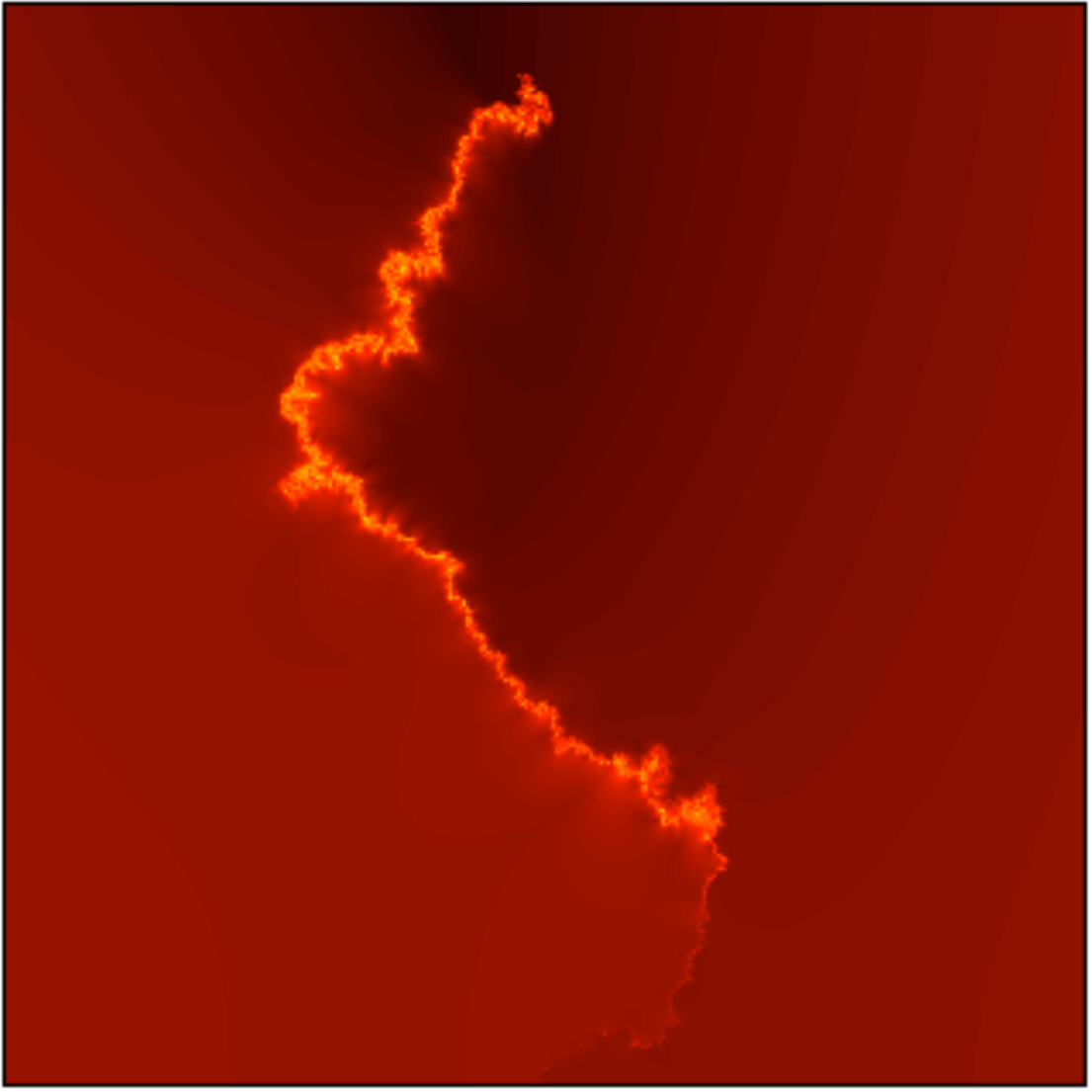}
}
\subfigure[$q=0.97$]{
\includegraphics[scale=0.35]{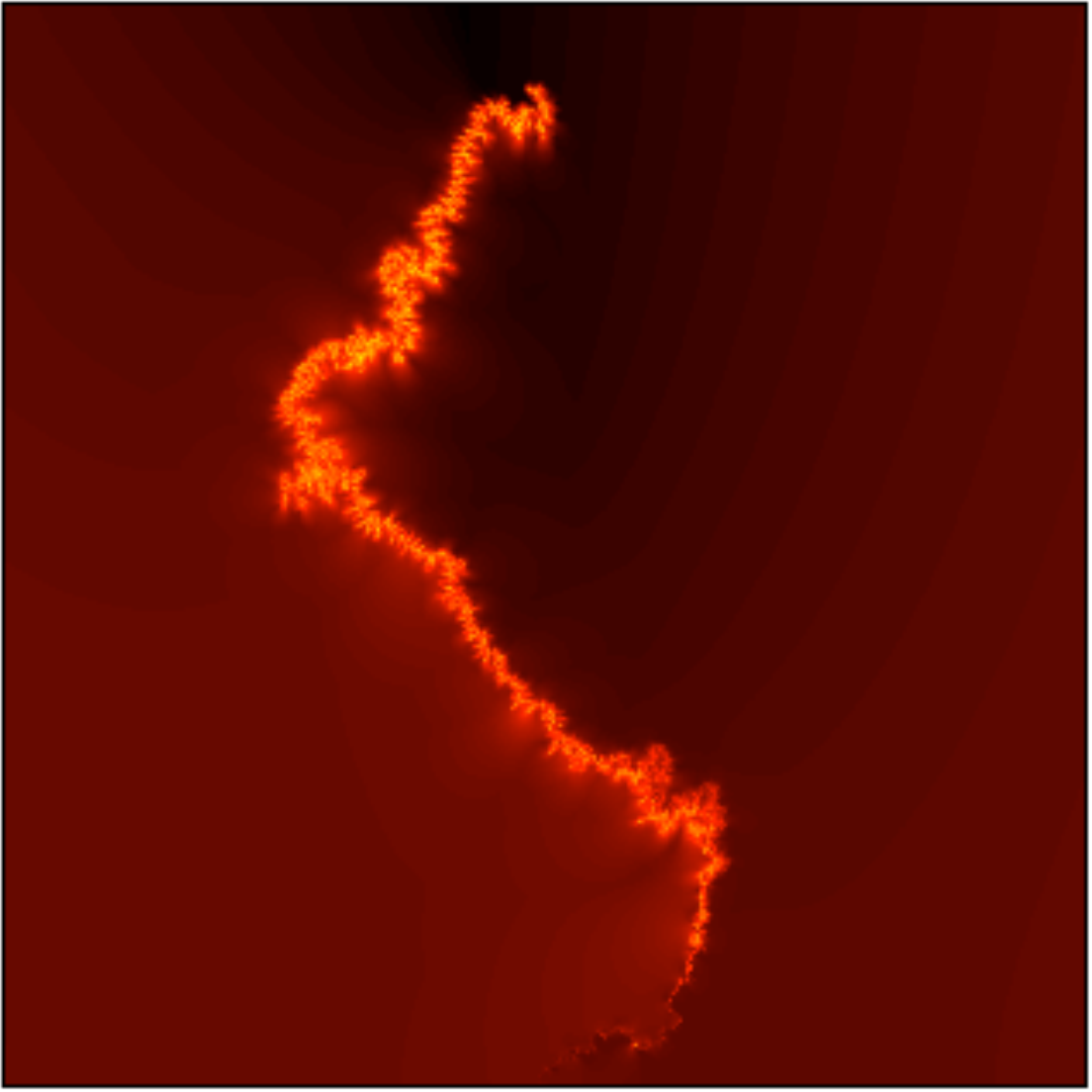}
}
\subfigure[$q=0.95$]{
\includegraphics[scale=0.35]{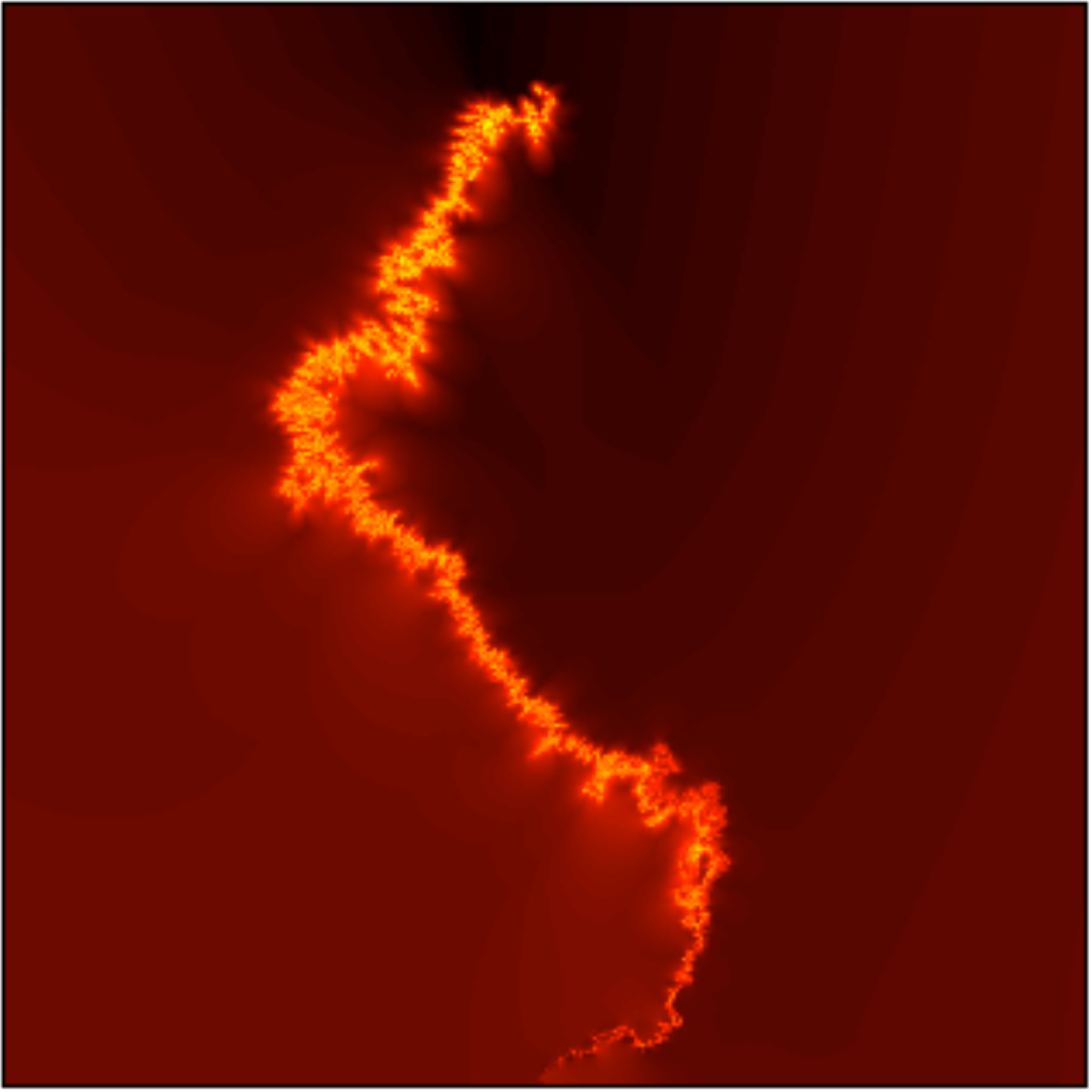}
}
\subfigure[$q=0.9$]{
\includegraphics[scale=0.35]{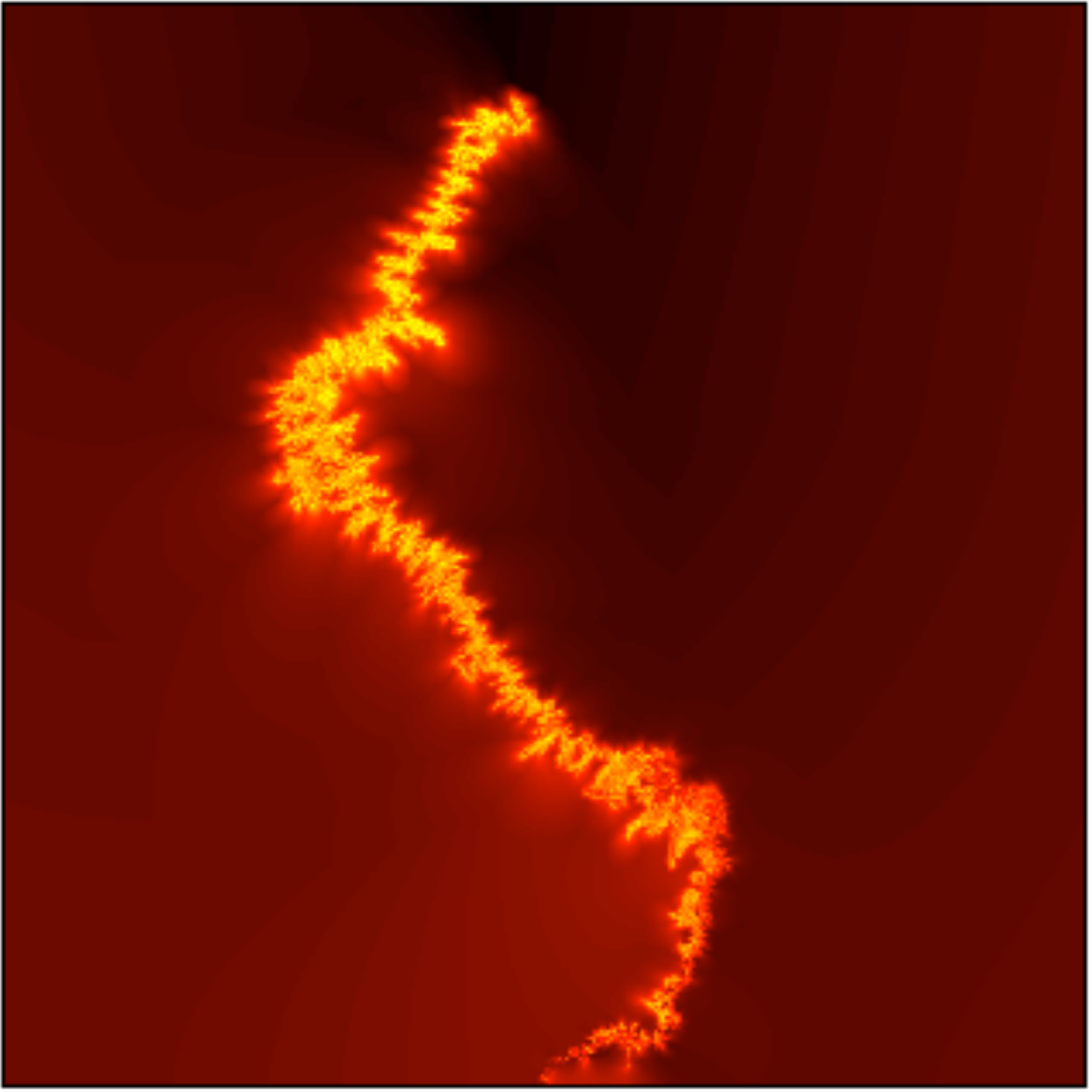}
}
\subfigure[$q=0.8$]{
\includegraphics[scale=0.35]{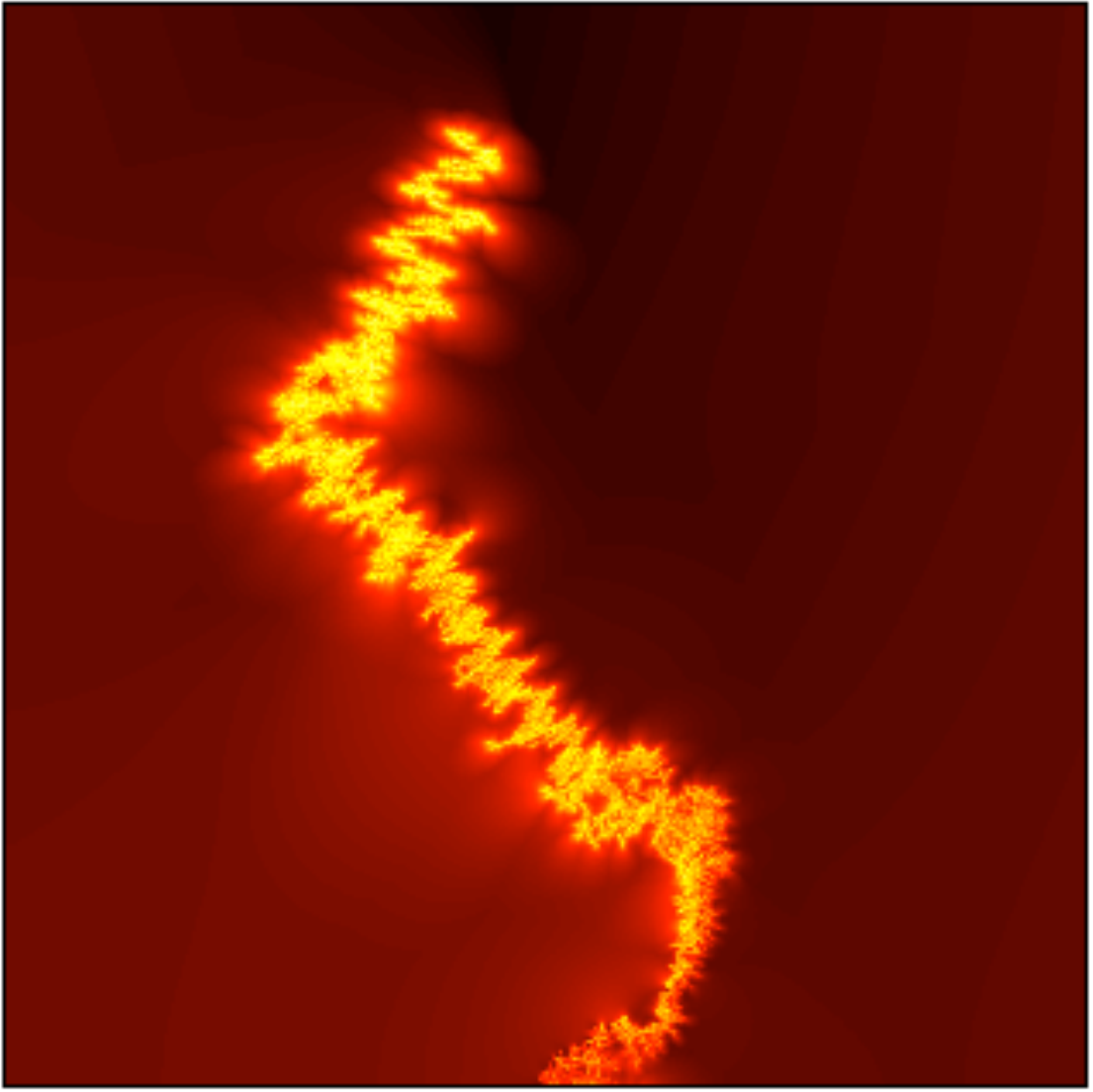}
}
\subfigure[$q=0.6$]{
\includegraphics[scale=0.35]{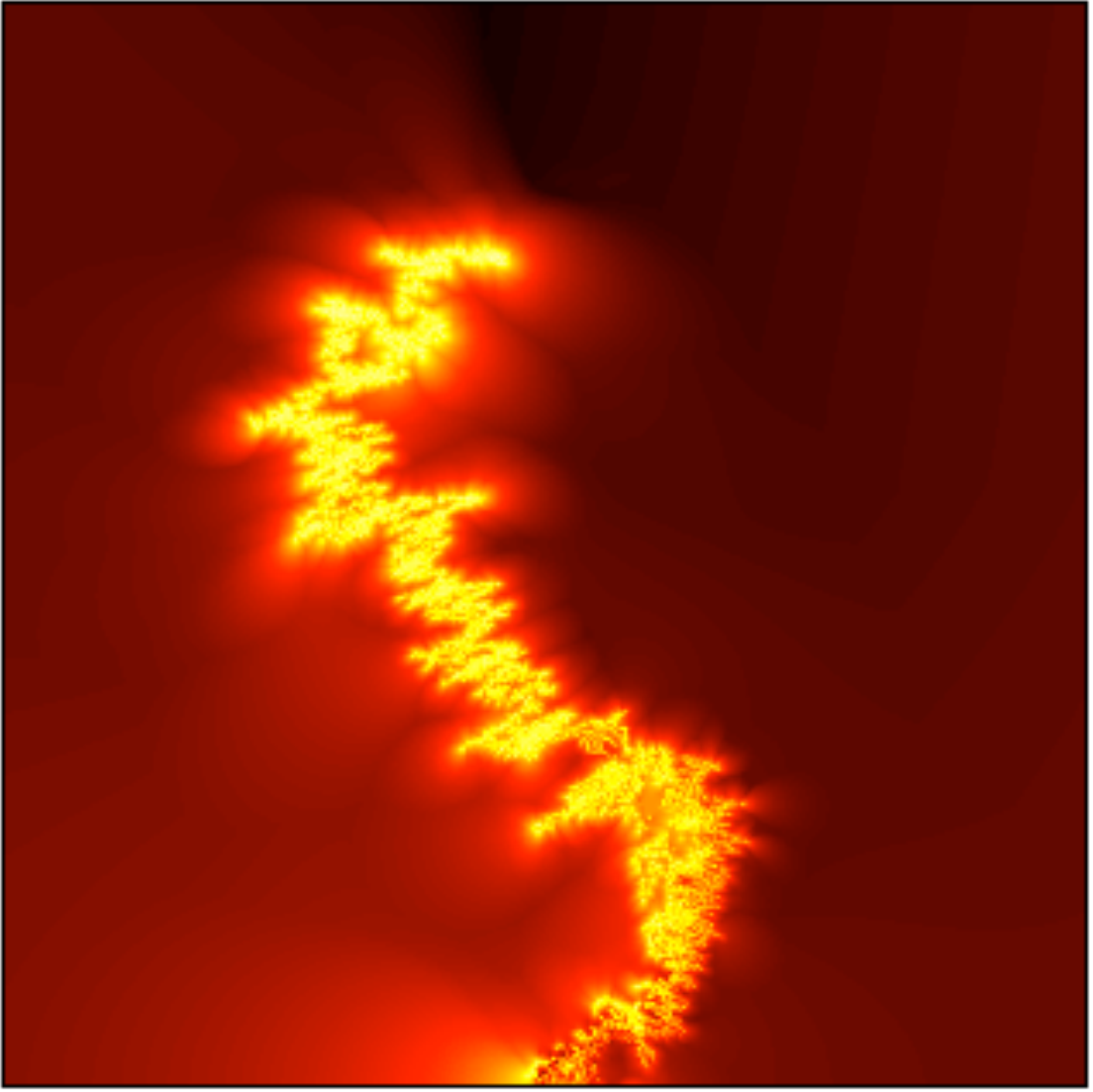}
}
\caption{
(Color online)
The hulls generated by Brownian motion $\sqrt{2}B_t$
display a ``dressing'' phenomenon, getting
thicker as $q$ decreases.
(The plots show $\delta_t(z)$ as in the previous figures.)
The same realization of $B_t$ was used for
all six different values of $q$.
}
\label{figure:brownian}
\end{figure}

A natural question is whether the different branches that are
seen with square-root driving can ever intersect, when one
introduces a non-monotonic force.
The answer is in the affirmative, as Fig.~\ref{figure:sin} shows
for the case of a sinusoidal driving function.
The branches, which are close together for high values of $q$,
wind around each other in a seemingly regular pattern,
and get further apart as $q$ decreases.
Finally, we show in Fig.~\ref{figure:brownian} how the stochastic version looks.
Here we have generated a typical realization of the brownian motion for $\kappa=2$
and produced plots of (\ref{eq:maxder}) for different values of $q$, keeping the
realization fixed.
As $q$ decreases from the classical value $1$, the curve gets dressed
with a thicker and thicker cloud of points, which are the stochastic
counterpart of the secondary curves observed for the sinusoidal and square-root
driving functions.
A deeper study of the measure on such objects is left for future work.

\section{Conlusions}

We have presented a generalization of the Loewner equation based on $q$-calculus;
it is defined essentially in the same way as the original equation,
with the ordinary time derivative replaced by a $q$-derivative.
The resulting equation is a finite-difference functional equation, whose
non-local nature permits the appearance of multiple curves, contrary to the classical
case, where a smooth enough driving function generates a simple curve.
(The hull generated by the evolution is defined in the usual way, as
the set of points for which the equation has become singular
before the present time.)
The explicit solution for the simplest case (constant driving function)
can be given in terms of well-known $q$-analogues of ordinary functions;
we call it the $q$-slit. It represents a single slit growing in the half plane, just
as in the classical case. Nonetheless, the form of the $q$-slit solution
already highlights some of the peculiar properties of the $q$-deformed
universe. It is not holomorphic on the half plane; it has an infinite number
of poles and zeros accumulating in the origin, where an essential singularity lies.
The complexity of the zero-pole structure increases when one considers
a square-root driving function, which would generate a tilted line in the
classical case. Here, it generates an infinite number of lines, all radiating
from $z=0$; the map has an infinite number of poles on each line, and
an infinite number of zeros on an infinite subset of lines.
The way in which this picture converges to the classical one when $q\to 1$
is an interesting realization of how $q$-deformation works,
and is an example of the richness one encounters when departing
from the common $q=1$ world.

We have proposed a numerical algorithm for approximating and simulating
this equation, which could be translated straightforwardly to other classes of $q$-differential equations.
We plan to employ it in the study of the stochastic $q$-deformed
equation, where the driving function is Brownian motion.
The main motivation for applying quantum calculus to the Loewner equation
is related to the subtle properties of $q$-derivatives: they cause the breaking
of the Markov property without compromising scale invariance.
The $q$-deformed SLE is expected to have fractal properties
dependent on both parameters $\kappa$ and $q$.

\begin{acknowledgements}
We gratefully acknowledge support from Universit\`a degli Studi di Milano
and Alessandro Vicini for GPU computing facilities.
We wish to thank Pietro Rotondo for stimulating discussions
and Sergio Caracciolo for reading the manuscript.
Both authors acknowledge financial support from Fondo Sociale Europeo (Regione Lombardia),
through the grant ``Dote Ricerca.''
\end{acknowledgements}

\end{document}